\documentclass[aps,prl,twocolumn,superscriptaddress]{revtex4-1}

\usepackage{graphicx}
\usepackage{amsmath}
\usepackage{amssymb}
\usepackage{wasysym}
\usepackage{textcomp}
\usepackage{color}

\definecolor{lila}{rgb}{0.5,0,1}
\newcommand{\bnen}{\begin{equation}}
\newcommand{\eden}{\end{equation}}
\newcommand{\bean}{\begin{eqnarray}}
\newcommand{\eean}{\end{eqnarray}}

\newcommand{\bna}{\begin{array}}
\newcommand{\eda}{\end{array}}

\begin{document}

\title{Dynamical topological quantum phase transitions in nonintegrable models}

\author{I. Hagym\'asi}
\affiliation{Department of Physics,
Arnold Sommerfeld Center for Theoretical Physics (ASC),
Fakult\"{a}t f\"{u}r Physik, Ludwig-Maximilians-Universit\"{a}t M\"{u}nchen,
D-80333 M\"{u}nchen, Germany}
\affiliation{Munich Center for Quantum Science and Technology (MCQST), Schellingstr. 4, D-80799 M\"unchen, Germany}
\affiliation{Strongly Correlated Systems "Lend\"ulet" Research Group, Institute for Solid State
Physics and Optics, MTA Wigner Research Centre for Physics, Budapest H-1525 P.O. Box 49, Hungary
}

\author{C. Hubig}
\affiliation{Max-Planck-Institut f\"ur Quantenoptik,
Hans-Kopfermann-Strasse 1, 85748 Garching, Germany}

\author{\"O. Legeza}
\affiliation{Strongly Correlated Systems "Lend\"ulet" Research Group, Institute for Solid State
Physics and Optics, MTA Wigner Research Centre for Physics, Budapest H-1525 P.O. Box 49, Hungary
}

\author{U. Schollw\"ock}
\affiliation{Department of Physics,
Arnold Sommerfeld Center for Theoretical Physics (ASC),
Fakult\"{a}t f\"{u}r Physik, Ludwig-Maximilians-Universit\"{a}t M\"{u}nchen,
D-80333 M\"{u}nchen, Germany}
\affiliation{Munich Center for Quantum Science and Technology (MCQST), Schellingstr. 4, D-80799 M\"unchen, Germany}

\date{\today}

\begin{abstract}
We consider sudden quenches across quantum phase transitions in the $S=1$
 XXZ model starting from the Haldane phase. We demonstrate that dynamical phase 
transitions may occur during these quenches that are identified by 
nonanalyticities in the rate function for the return probability. In addition, 
we show that the temporal behavior of the string order parameter is 
intimately related to the subsequent dynamical phase transitions. We furthermore find that the dynamical 
quantum phase transitions can be accompanied by enhanced two-site entanglement. 
\end{abstract}

\maketitle
\emph{Introduction.---} 
Nonequilibrium dynamics of many-body quantum systems under unitary time 
evolution continue to pose a
challenging problem. The time evolution of a quantum system after a sudden global quench plays a 
distinguished role in this field since this process can be routinely carried out in experiments and
it is addressable in theoretical calculations \cite{Essler_2016}. 
\par The quench process is even more interesting when it drives the system through an equilibrium 
phase 
transition. This has opened up a new area of research named dynamical quantum 
phase transitions 
(DQPTs) \cite{PhysRevLett.115.140602,Heyl_2018,PhysRevLett.110.135704}.
Although they are not in one-to-one correspondence with the equilibrium phase 
transitions, but rather a 
new form of 
critical behavior, they often emerge when the quench crosses a phase 
transition. Recently, direct experimental observation of DQPTs has been 
reported, where a transverse-field Ising model was realized with trapped 
ions \cite{PhysRevLett.119.080501,Weitenberg:nature}.
For the better 
understanding of 
DQPTs several integrable models have been 
considered \cite{PhysRevLett.110.135704,PhysRevB.93.085416,PhysRevB.91.155127, PhysRevB.89.161105,PhysRevB.97.174401,PhysRevLett.121.130603}, where the time evolution can be solved 
exactly. It has 
been revealed that, like equilibrium phase transitions, DQPTs also affect other observables. For 
example, when 
the quench starts from a broken-symmetry phase, where the order can be characterized by a local 
order 
parameter, the order parameter exhibits a temporal decay with a series of times 
where it vanishes \cite{PhysRevLett.110.135704,PhysRevB.96.134313}. 
These times 
usually coincide with the 
times where the DQPTs occur. The case is more difficult when one considers a nonintegrable model \cite{PhysRevB.96.134427,PhysRevE.96.062118,PhysRevB.96.104436,Halimeh:preprint1,Halimeh:preprint2,Halimeh:preprint3}. 
Namely, 
the obvious choice of an observable (e.g.~the equilibrium order parameter) may 
not follow the dynamics dictated by the DQPTs and the connection between them remains 
elusive like in 
the case of a nonintegrable Ising chain \cite{PhysRevB.87.195104} 
or Bose-Hubbard model \cite{Fogarty_2017}. Thus, the relation of DQPTs to 
observables in nonintegrable models deserves further investigation in general.
\par In this work we study quenches starting from the Haldane phase to 
regimes where the ground state has trivial topology. The Haldane phase is a 
paradigmatic example of a symmetry-protected topological phase, and has direct 
relevance in quantum 
operations \cite{PhysRevA.82.012328,PhysRevLett.108.240505}. The 
understanding of topological phases under unitary time evolution hence is of 
particular interest \cite{PhysRevB.99.075148}. More precisely we
examine the time evolution of the string order parameter (SOP) in 
the $S=1$ XXZ 
model.
Although this model has been investigated before regarding the 
thermalization of string 
order \cite{PhysRevB.94.024302,PhysRevB.90.020301}, we 
point out that its dynamics is much richer and there is 
a so far unrecognized connection between the dynamics of the SOP and the 
underlying 
DQPTs, which 
have also not been observed before. 
More precisely, we consider several types of quenches and demonstrate that if a DQPT occurs during 
the time 
evolution, then 
it is accompanied by the zero of one of the SOPs. On the other hand, if DQPTs 
are not 
present, then all the three SOPs exhibit a smooth decay without any zeros. Such a link has been reported for a \emph{noninteracting} 
system \cite{PhysRevB.93.085416}. Our analysis suggests that 
this 
correspondence is not only a property of exactly solvable systems, but appears to be valid on a 
more 
general level. Moreover, our findings also reveal that the dynamics of the 
SOP is not only influenced by the symmetry of the quench 
Hamiltonian \cite{PhysRevB.90.020301,PhysRevLett.121.090401}, but by the 
crossing of a phase boundary as well. This conclusion
 is supported by a quantum information analysis of the time-evolved
 wave function, where the DQPT manifests itself in the enhancement of
 the two-site entanglement. 
\par\emph{Model and methods.---}
The XXZ Heisenberg model can be written as follows: 
\begin{equation}
\mathcal{H}=\sum_{i=1}^{L-1} [J(S^x_iS^x_{i+1}+S^y_iS^y_{i+1}) + \Delta S^z_iS^z_{i+1} ] + D\sum_{i=1}^L (S_i^z)^2,
\end{equation}
where $S_i^{\alpha}$ denotes the appropriate spin-1 operator component. 
The 
parameters $\Delta$ and $D$ denote the Ising and uniaxial single-ion 
anisotropy, respectively and we set $J=1$ to define the energy scale. We also set $\hbar=1$, thus, 
the time is measured in units of $1/J$.  The ground state of this model \cite{PhysRevB.28.3914} has been thoroughly 
investigated in the past few decades, and its phase diagram is now 
well-known \cite{PhysRevB.67.104401}. Besides the symmetry-protected Haldane phase, for  
$D/J\gg 1$ it realizes a trivial singlet phase, for $\Delta/J\gg 1$ a N\'eel-like ground state, while for  
$\Delta/J\ll -1$ a ferromagnetic ground state occurs. Between the ferromagnetic and Haldane phases, 
a critical XY regime turns up. Moreover, various types of phase transitions take place at the phase boundaries allowing us to examine their dynamical counterparts within the framework of a single model.
In what follows we study quenches across quantum phase transitions originating from the Haldane phase.
More precisely, we initialize the system in the ground-state of the 
Affleck-Kennedy-Lieb-Tasaki (AKLT) Hamiltonian \cite{PhysRevLett.59.799,AKLT}, 
$|\Psi_0\rangle$, with $\langle\Psi_0|\sum_iS_i^z|\Psi_0\rangle=0$, then we let 
it evolve unitarily 
governed by the Hamiltonian 
$\mathcal{H}$,
$|\Psi(t)\rangle=e^{-i\mathcal{H}t}|\Psi_0\rangle.$ 
The time evolution \cite{SI} is carried out using the 
time-dependent 
variational principle
method \cite{PhysRevLett.107.070601,PhysRevB.94.165116,Hubig:review}.
\par\emph{Quench results.---}
Because we are not close to equilibrium by any means, 
the steady 
state is not expected to exhibit the same properties as the corresponding ground state of the 
postquench 
Hamiltonian. The positions of the phase transitions should only serve as guides when to expect the 
emergence 
of dynamical criticality. DQPTs are difficult to characterize in contrast to equilibrium phase 
transitions due to the 
lack of a free energy. Nevertheless, it has been shown by several works that they manifest themselves as 
nonanalyticities in 
the rate function of the return 
probability \cite{PhysRevLett.115.140602,Heyl_2018,PhysRevLett.110.135704,PIROLI2018454}. The 
return probability called the Loschmidt 
echo 
$\mathcal{L}(t)$ is defined 
as: $\mathcal{L}(t)=|\langle\Psi_0|\Psi(t)\rangle|^2$.
 Since $\mathcal{L}(t)$  is not well-defined in the thermodynamic limit, it is therefore convenient 
to introduce the rate function, $\lambda(t)$, which reads
$\lambda(t)=-\frac{1}{L}\log[\mathcal{L}(t)]$.
The rate function can be regarded as a dynamical analogue of the free energy density, and just like 
the free 
energy at equilibrium phase transitions, this quantity can exhibit nonanalytic behavior at critical 
times as well.
In one-dimensional systems the nonanalyticities show up as kinks during the time evolution.
\par
\begin{figure}[!t]
\includegraphics[scale=0.6]{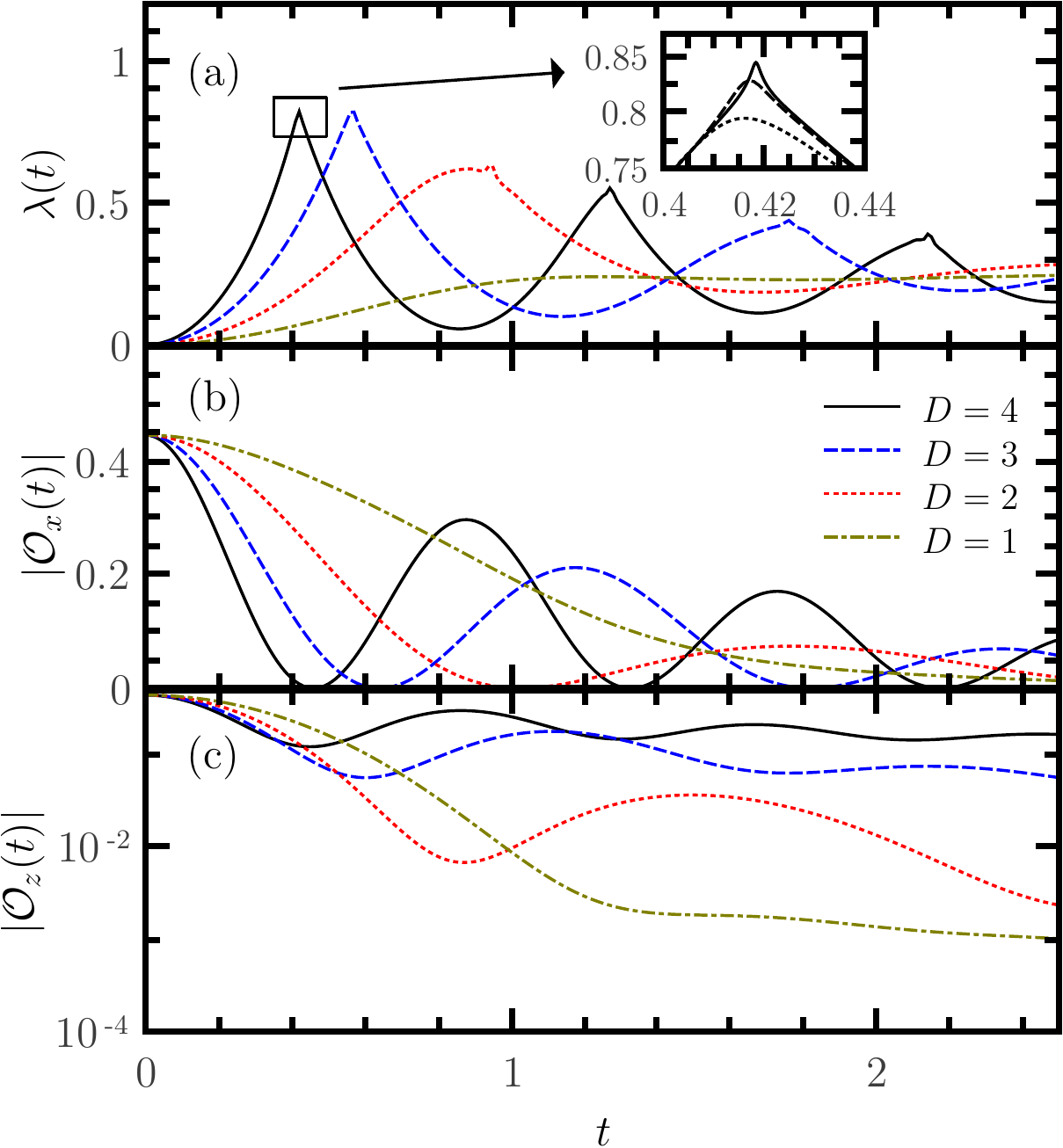}
\caption{(a) The main panel shows the rate function for $L=80$, $\Delta=1$ and various values of 
$D$ as 
indicated by the legend in panel (b). The inset figure of panel (a) shows the finite-size effects 
around the first 
critical time for $D=4$, the dotted, dashed and solid lines denote $L=30,60$ and 120, respectively. 
(b) The 
$x$-component of the SOP for various values of $D$, $\Delta=1$ and for chain 
length $L=80$. (c) 
The $z$-component of the SOP for the same parameters as in panels (a) and (b).}
\label{fig:quench-D}
\end{figure}

\begin{figure}[!t]
\includegraphics[scale=0.7]{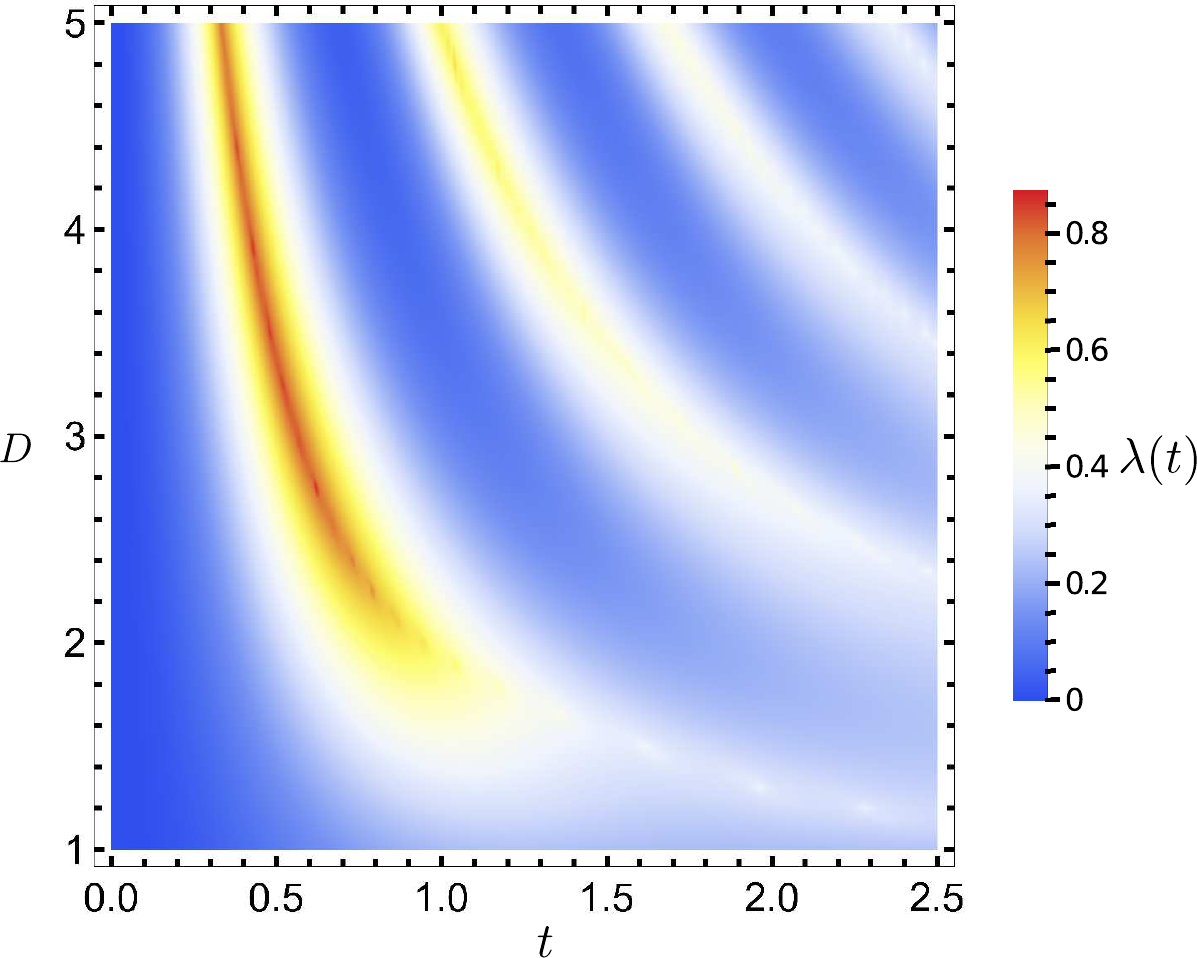}
\caption{The rate function as a function of time and single-ion anisotropy using 
a color code in the sidebar, for $\Delta=1$ and $L=80$. }
\label{fig:quench-D-full}
\end{figure}

 The other important quantity of interest is the string operator, which is defined as
\begin{equation}
\label{eq:string-op}
\hat{\mathcal{O}}^{\alpha}_{\ell}= \hat{S}_j^{\alpha} \left[ \prod_{n=j+1}^{j+\ell-1}e^{i\pi \hat{S}
_n^{\alpha}} \right] \hat{S}_{j+\ell}^{\alpha}  \quad (\alpha\in \{x,y,z\}).
\end{equation}
With the help of these operators, the hidden topological order within the Haldane phase can be 
characterized by nonlocal order 
parameters \cite{PhysRevB.40.4709,Pollmann2010prb,
Pollmann2012prb}, that is
$\mathcal{O}^{\alpha}=\lim_{\ell\rightarrow\infty}\left\langle 
\hat{\mathcal{O}}^{\alpha}_{\ell} \right\rangle \neq 0$
should hold for any $\alpha$, where the expectation value is taken with respect to the ground state.
 In case of the AKLT state its value is exactly 
$\mathcal{O}^{\alpha}=-4/9$ \cite{Kl_mper_1993}. In what follows we define string order in the time-evolved state similarly, namely:
$\mathcal{O}^{\alpha}(t)=\lim_{\ell\rightarrow\infty}\langle\Psi(t)|
\hat{\mathcal{O}}^{\alpha}_{\ell}|\Psi(t) \rangle.$
We study quenches into the large-$D$ phase first. The numerical results are summarized in 
Figs.~\ref{fig:quench-D} and \ref{fig:quench-D-full}.
We can easily see from Fig.~\ref{fig:quench-D}(a) and 
Fig.~\ref{fig:quench-D-full} that the rate function displays distinct 
behavior 
depending on the value of $D$. For $D=1$, the system is quenched in the vicinity of the phase 
transition point 
between the Haldane and the large-$D$ phase, in spite of that, it shows completely analytic, smooth 
behavior. 
Accordingly, both $\mathcal{O}^x(t)$ and $\mathcal{O}^z(t)$ decay monotonically without any zeros. 
For 
$D=2$, the quench drives the the system through the phase transition, and around $t\approx 1$ a 
tiny kink 
appears in the rate function. At the same time the $\mathcal{O}^x(t)$ becomes 
zero, while 
$\mathcal{O}^z(t)$ 
remains finite. For even larger values of $D$ we can observe well-developed kinks in the rate 
functions, and 
they appear more frequently as the single-ion anisotropy is increased. Not only can we conclude that 
the 
difference between the positions of the kinks agree very well with the positions 
of the zeros of 
$\mathcal{O}^x(t)
$, but the individual kink positions agree also well with the zeros of $\mathcal{O}^x(t)$. We can 
also notice that the regions where $\mathcal{O}^x(t)$ shows local maxima coincide with the local 
minima of the rate function.
In the inset of 
Fig.~\ref{fig:quench-D}(a) we can observe the finite-size dependence of the rate function around 
the critical 
times. We can indeed identify that a kink is being developed as the system size is increased, thus, 
for infinite 
system size, the time-evolved state becomes completely orthogonal to the initial state inducing a 
DQPT in the system. At these critical times, the system should possess trivial topology, since 
$\mathcal{O}^x(t)$ 
vanishes here. It is also 
worth mentioning that $\mathcal{O}^z(t)$ decays smoothly with some superimposed oscillations for 
larger $D$, but it does not show any zeros in this time window. Interestingly, it decays slower in 
time for larger values of $D$. This seems to be analogous to what happens during an interaction quench 
in the Fermi-Hubbard model. Namely, when the quench drives the system from a weakly interacting regime to a 
strongly interacting one, then the relaxation time of the double occupancy increases exponentially with the 
Hubbard-$U$ \cite{PhysRevB.84.035122,PhysRevB.86.205101}. In our case, the nonzero spin components play
the same role, when we quench to the large-$D$ phase. 
\par Next, we turn our attention to quenches into the N\'eel regime.
\begin{figure}[!t]
\includegraphics[scale=0.6]{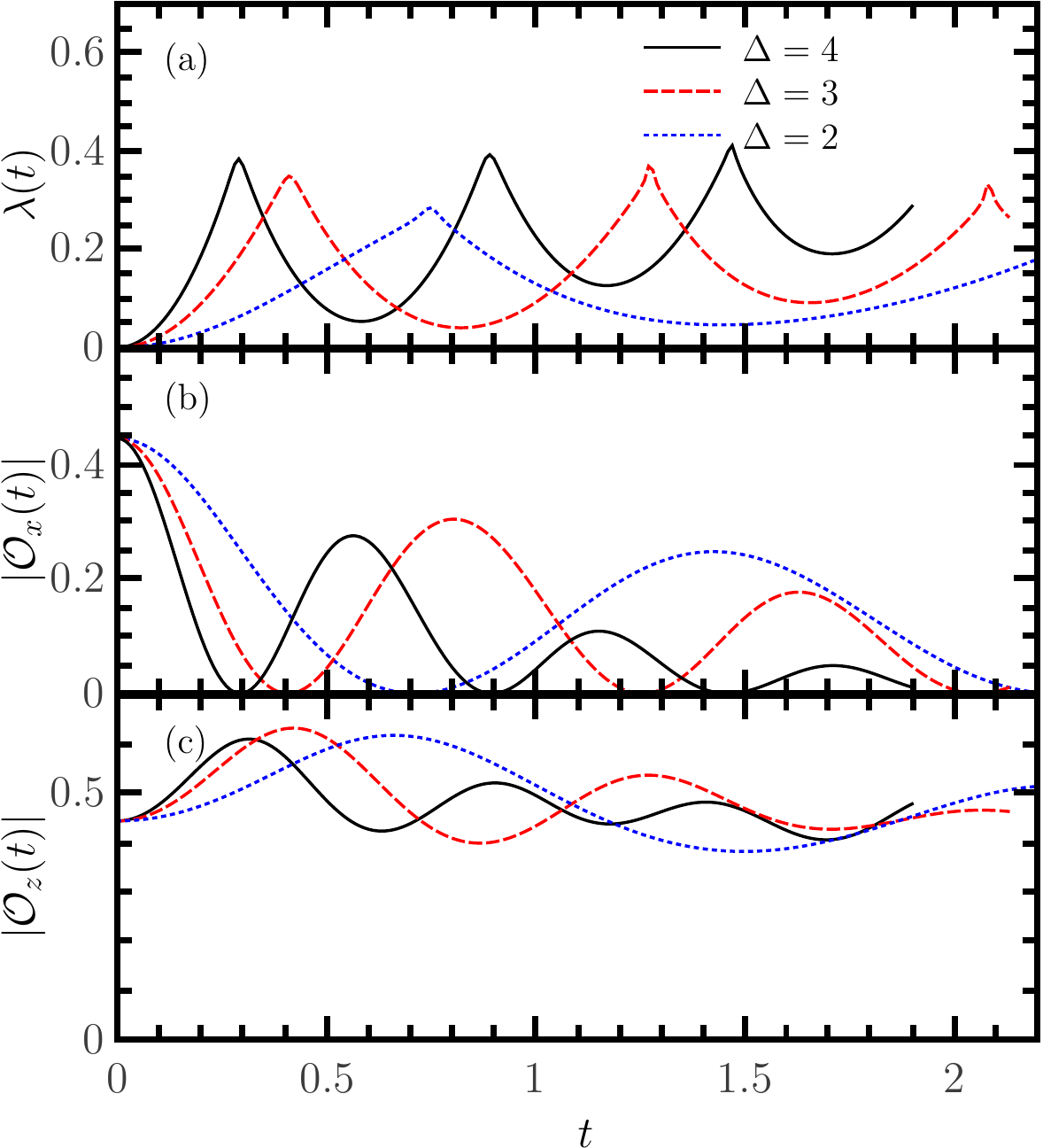}
\caption{(a) The rate function for various values of $\Delta$ and $D=0$. Panels (b) and (c) show 
the 
$x$- and $z$-component of the SOP for the same values of $\Delta$ and $D=0$ as 
in panel (a).  The chain length is $L=80$ in all cases.}
\label{fig:quench-Delta}
\end{figure}
In that case already for $\Delta=2$ well-developed kinks are observable. Similarly, the zeros of 
the 
$\mathcal{O}^x(t)$ are in excellent agreement with the kinks in the rate 
function. In addition, 
$\mathcal{O}^z(t)$
remains always nonzero, moreover, its value is slightly increased after the quench. This may not 
surprise us,
if we recall that the Ising term with $\Delta>1$ enhances the antiferromagnetic correlations 
leading therefore
to an increased string correlation value in the $z$ direction.
\par We also consider quenches with $\Delta<0$ \cite{SI}, which corresponds to probing the spectrum of the 
XY and the ferromagnetic phase. In these cases we find no signs of DQPTs in the rate function
and the SOP decays without any zeros in time. It is worth noting that the relation between the DQPTs and the 
SOP holds even if the model has fewer symmetries \cite{SI} or in the full SU(2)-symmetric bilinear-biquadratic chain for the Haldane--dimerized transition.
\par \emph{Discussion.---}
Having seen how the rate function and the SOPs behave in the different cases, 
the question
naturally arises if we can say something more to account for the different 
behavior in the 
$\Delta<0$ cases. Our 
only guide is the phase diagram \cite{PhysRevB.67.104401}.  When we quench to the N\'eel or the 
large-$D$ phase, we face in
both cases a second-order phase transition. More precisely, the Haldane--large-$D$ critical line is 
a Gaussian-, while the Haldane--N\'eel one is an Ising-type phase 
transition \cite{PhysRevB.67.104401}. During these quenches we cross the 
critical lines,
where the spectrum is expected to be gapless. Thus, this drastic change manifests itself in the 
appearance of 
DQPTs in both cases.  In contrast, when we quench to the XY phase, an infinite-order phase transition 
takes place 
at the phase boundary \cite{PhysRevB.67.104401}. Since the gap opens here 
exponentially slowly, it may prevent the occurrence of 
DQPTs. 
\begin{figure*}[!ht]
\includegraphics[scale=0.45]{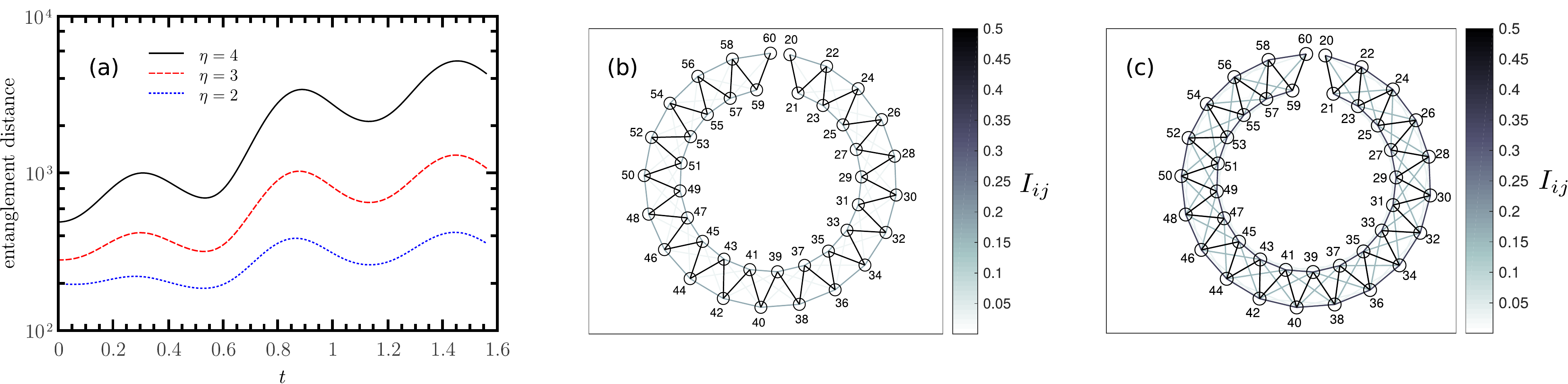} \
\caption{(a) The entanglement distance as a function of time calculated with different exponents 
for the quench with $\Delta=4$, $D=0$. (b) The entanglement patterns in the chain at time $t=0.53$ for the same quench as in panel (a). The lines encode the magnitude of the mutual information between 
different sites using the grayscale sidebar. Note that 20 sites are discarded at both ends. (c) Similar to panel (b) but for $t=0.89$. The chain length is $L=80$ in all cases.}
\label{fig:ent-patterns}
\end{figure*}
During the quench from the Haldane phase to the ferromagnetic phase, two phase transitions are encountered,
infinite order (Haldane to XY) and first order (XY to ferromagnetic) \cite{PhysRevB.67.104401}. The lack of DQPTs in 
this case
can be traced back to several reasons. A first-order transition involves a level crossing of the 
ground state with a higher-lying state at the transition point, but the whole spectrum does not 
change so drastically at the transition point like in the case of a second-order transition. 
Another reason, which could also apply for the quench in the XY regime, is that DQPTs may occur at 
a later time and simply our simulation time is not long enough. 
\par \emph{Quantum information analysis.---}
In the analysis of equilibrium quantum phase transitions the tools of the quantum information theory 
turned out
to be extremely 
useful \cite{legeza2003b,vidallatorre03,calabrese04,rissler2006,legeza2006,
luigi2008}.  The entropies of various subsystems can be sensitive indicators for phase transitions but also 
for characterizing 
the entanglement structure of the wave 
function \cite{gu:prl2004,wu:prl2004,yang:pra2005,deng:prb2006}, 
especially through the mutual information, $I_{ij}$, defined as
$I_{ij}=s_i+s_j-s_{ij},$
which measures all correlations both of classical and quantum origin between sites $i$ and $j$. Here $s_i$ and 
$s_{ij}$ denote the one- and two-site entropies of the corresponding sites.
One can naturally ask if there appear some anomalous signatures in these quantities around the 
critical times. 
The entanglement entropy in global quenches, like ours, increases linearly with the time, thus 
complicating the 
analysis. It has been reported before that the DQPTs may result in enhanced entropy 
production 
around the critical times \cite{PhysRevLett.119.080501}. We ask the question if DQPTs affect the two-site entanglement, since it 
is easier to 
interpret than the whole entanglement entropy, which contains the cumulated effects of several 
physical 
processes. To this end, we adopt the entanglement 
distance \cite{PhysRevA.83.012508,RISSLER2006519}, $I_{\rm dist}^{(\eta)}$, to 
quantify the 
strength of the two-site entanglement in the system:
$I_{\rm dist}^{(\eta)}=\sum_{ij}I_{ij}|i-j|^{\eta}.$
Different $\eta$ exponents have been used in the literature. For negative 
(positive) values of 
$\eta$ it emphasizes the contribution of short-range (long-range) 
entanglement \cite{PhysRevA.83.012508,RISSLER2006519}.
We calculate this quantity using different exponents to study the delocalization of the entanglement
 during the 
time evolution. This is shown in Fig.~\ref{fig:ent-patterns}(a) in the case when the quench goes through 
the Haldane--N\'eel transition line. Remarkably, $I_{\rm dist}^{(\eta)}$ does not increase monotonically as one 
would naively 
expect from the linear increase of the entanglement entropy (not shown).  More surprising is the 
fact that the 
positions of the maxima in $I_{\rm dist}^{(\eta)}$ agree very well with the positions of the DQPTs. 
We can 
interpret this phenomenon such that the two-site correlations get enhanced around the critical 
times, resembling 
to the long-range correlations occurring at the equilibrium phase transition. It is important to 
note, however, 
that truly long-range correlations cannot occur in our case, since the Lieb-Robinson bound does not 
allow 
the instantaneous buildup of long-ranged 
correlations \cite{RevModPhys.36.856,Lieb1972}. We illustrate this enhancement 
in 
Fig.~\ref{fig:ent-patterns}(b) and (c), where the two times correspond to a local minimum 
and maximum in $I_{\rm dist}^{(\eta)}$, respectively.
\par We also checked if a similar connection exists for the quench through the Haldane--large-$D$ 
line. Although
we found that $I_{\rm dist}^{(\eta)}$ is also a nonmonotonic function of time, its relation to the 
DQPTs is less clear. This discrepancy may be understood with the help of the following
argument: the quantum critical point, in a strict sense, shows up only at zero temperature, the critical 
fluctuations, however, influence the finite-temperature properties as well, which can be detected in a wide range
of temperatures. In our case, the quench energy can be regarded as an effective temperature, and by crossing a 
a quantum critical point, we probe the critical region 'above' the critical point. 
Since the universality classes of the two phase transitions 
are distinct, the associated critical regions are also expected to display different behavior.
\par \emph{Conclusions.---}
We examined the nonequilibrium dynamics of the Haldane phase under 
unitary 
time evolution in the XXZ model. 
We revealed that DQPTs can occur when the quench crosses a phase boundary and they manifest themselves in
nonanalyticities in the rate function for the Loschmidt echo. 
Moreover, we demonstrated that 
there is an intrinsic
connection between the nonanalyticities and the zeros of the SOP.  Thus, the 
emerging 
nonequilibrium time scale has also fingerprints in other observable quantities.
Using the tools of quantum information theory 
we pointed out that the two-site entanglement may get significantly 
enhanced in the vicinity of the DQPTs exhibiting some resemblance to equilibrium phase transitions.
Since both the Loschmidt echo and SOP are now 
within the reach of experimental 
techniques \cite{PhysRevLett.119.080501,Hilker484}, our findings could be also 
directly tested in the future. In recent experiments effective spin-1 chains 
have already been successfully simulated with trapped 
ions \cite{PhysRevX.5.021026}.
\par\emph{Acknowledgments.---}We acknowledge discussions with N.~Cooper and M.~McGinley.  
I.H.~and \"O.L.~were supported by the Alexander 
von Humboldt Foundation and in part by Hungarian 
National Research,
Development and Innovation Office (NKFIH) through Grant No. K120569 and the Hungarian Quantum 
Technology National Excellence
Program (Project No.  2017-1.2.1-NKP-2017-00001). C.H.~acknowledges funding 
through ERC Grant QUENOCOBA, ERC-2016-ADG (Grant no. 742102). This work was 
also supported in part by the Deutsche Forschungsgemeinschaft (DFG, German 
Research
Foundation) under Germany’s Excellence Strategy -- EXC-2111 -- 390814868.

\bibliography{tki_refs.bib}

\begin{thebibliography}{58}%
\makeatletter
\providecommand \@ifxundefined [1]{%
 \@ifx{#1\undefined}
}%
\providecommand \@ifnum [1]{%
 \ifnum #1\expandafter \@firstoftwo
 \else \expandafter \@secondoftwo
 \fi
}%
\providecommand \@ifx [1]{%
 \ifx #1\expandafter \@firstoftwo
 \else \expandafter \@secondoftwo
 \fi
}%
\providecommand \natexlab [1]{#1}%
\providecommand \enquote  [1]{``#1''}%
\providecommand \bibnamefont  [1]{#1}%
\providecommand \bibfnamefont [1]{#1}%
\providecommand \citenamefont [1]{#1}%
\providecommand \href@noop [0]{\@secondoftwo}%
\providecommand \href [0]{\begingroup \@sanitize@url \@href}%
\providecommand \@href[1]{\@@startlink{#1}\@@href}%
\providecommand \@@href[1]{\endgroup#1\@@endlink}%
\providecommand \@sanitize@url [0]{\catcode `\\12\catcode `\$12\catcode
  `\&12\catcode `\#12\catcode `\^12\catcode `\_12\catcode `\%12\relax}%
\providecommand \@@startlink[1]{}%
\providecommand \@@endlink[0]{}%
\providecommand \url  [0]{\begingroup\@sanitize@url \@url }%
\providecommand \@url [1]{\endgroup\@href {#1}{\urlprefix }}%
\providecommand \urlprefix  [0]{URL }%
\providecommand \Eprint [0]{\href }%
\providecommand \doibase [0]{http://dx.doi.org/}%
\providecommand \selectlanguage [0]{\@gobble}%
\providecommand \bibinfo  [0]{\@secondoftwo}%
\providecommand \bibfield  [0]{\@secondoftwo}%
\providecommand \translation [1]{[#1]}%
\providecommand \BibitemOpen [0]{}%
\providecommand \bibitemStop [0]{}%
\providecommand \bibitemNoStop [0]{.\EOS\space}%
\providecommand \EOS [0]{\spacefactor3000\relax}%
\providecommand \BibitemShut  [1]{\csname bibitem#1\endcsname}%
\let\auto@bib@innerbib\@empty
%</preamble>
\bibitem [{\citenamefont {Essler}\ and\ \citenamefont
  {Fagotti}(2016)}]{Essler_2016}%
  \BibitemOpen
  \bibfield  {author} {\bibinfo {author} {\bibfnamefont {F.~H.~L.}\
  \bibnamefont {Essler}}\ and\ \bibinfo {author} {\bibfnamefont
  {M.}~\bibnamefont {Fagotti}},\ }\href {\doibase
  10.1088/1742-5468/2016/06/064002} {\bibfield  {journal} {\bibinfo  {journal}
  {J. Stat. Mech.: Theory and Exp.}\ }\textbf {\bibinfo {volume} {2016}},\
  \bibinfo {pages} {064002} (\bibinfo {year} {2016})}\BibitemShut {NoStop}%
\bibitem [{\citenamefont {Heyl}(2015)}]{PhysRevLett.115.140602}%
  \BibitemOpen
  \bibfield  {author} {\bibinfo {author} {\bibfnamefont {M.}~\bibnamefont
  {Heyl}},\ }\href {\doibase 10.1103/PhysRevLett.115.140602} {\bibfield
  {journal} {\bibinfo  {journal} {Phys. Rev. Lett.}\ }\textbf {\bibinfo
  {volume} {115}},\ \bibinfo {pages} {140602} (\bibinfo {year}
  {2015})}\BibitemShut {NoStop}%
\bibitem [{\citenamefont {Heyl}(2018)}]{Heyl_2018}%
  \BibitemOpen
  \bibfield  {author} {\bibinfo {author} {\bibfnamefont {M.}~\bibnamefont
  {Heyl}},\ }\href {\doibase 10.1088/1361-6633/aaaf9a} {\bibfield  {journal}
  {\bibinfo  {journal} {Rep. Prog. Phys.}\ }\textbf {\bibinfo {volume} {81}},\
  \bibinfo {pages} {054001} (\bibinfo {year} {2018})}\BibitemShut {NoStop}%
\bibitem [{\citenamefont {Heyl}\ \emph {et~al.}(2013)\citenamefont {Heyl},
  \citenamefont {Polkovnikov},\ and\ \citenamefont
  {Kehrein}}]{PhysRevLett.110.135704}%
  \BibitemOpen
  \bibfield  {author} {\bibinfo {author} {\bibfnamefont {M.}~\bibnamefont
  {Heyl}}, \bibinfo {author} {\bibfnamefont {A.}~\bibnamefont {Polkovnikov}}, \
  and\ \bibinfo {author} {\bibfnamefont {S.}~\bibnamefont {Kehrein}},\ }\href
  {\doibase 10.1103/PhysRevLett.110.135704} {\bibfield  {journal} {\bibinfo
  {journal} {Phys. Rev. Lett.}\ }\textbf {\bibinfo {volume} {110}},\ \bibinfo
  {pages} {135704} (\bibinfo {year} {2013})}\BibitemShut {NoStop}%
\bibitem [{\citenamefont {Jurcevic}\ \emph {et~al.}(2017)\citenamefont
  {Jurcevic}, \citenamefont {Shen}, \citenamefont {Hauke}, \citenamefont
  {Maier}, \citenamefont {Brydges}, \citenamefont {Hempel}, \citenamefont
  {Lanyon}, \citenamefont {Heyl}, \citenamefont {Blatt},\ and\ \citenamefont
  {Roos}}]{PhysRevLett.119.080501}%
  \BibitemOpen
  \bibfield  {author} {\bibinfo {author} {\bibfnamefont {P.}~\bibnamefont
  {Jurcevic}}, \bibinfo {author} {\bibfnamefont {H.}~\bibnamefont {Shen}},
  \bibinfo {author} {\bibfnamefont {P.}~\bibnamefont {Hauke}}, \bibinfo
  {author} {\bibfnamefont {C.}~\bibnamefont {Maier}}, \bibinfo {author}
  {\bibfnamefont {T.}~\bibnamefont {Brydges}}, \bibinfo {author} {\bibfnamefont
  {C.}~\bibnamefont {Hempel}}, \bibinfo {author} {\bibfnamefont {B.~P.}\
  \bibnamefont {Lanyon}}, \bibinfo {author} {\bibfnamefont {M.}~\bibnamefont
  {Heyl}}, \bibinfo {author} {\bibfnamefont {R.}~\bibnamefont {Blatt}}, \ and\
  \bibinfo {author} {\bibfnamefont {C.~F.}\ \bibnamefont {Roos}},\ }\href
  {\doibase 10.1103/PhysRevLett.119.080501} {\bibfield  {journal} {\bibinfo
  {journal} {Phys. Rev. Lett.}\ }\textbf {\bibinfo {volume} {119}},\ \bibinfo
  {pages} {080501} (\bibinfo {year} {2017})}\BibitemShut {NoStop}%
\bibitem [{\citenamefont {Fl{\"a}schner}\ \emph {et~al.}(2018)\citenamefont
  {Fl{\"a}schner}, \citenamefont {Vogel}, \citenamefont {Tarnowski},
  \citenamefont {Rem}, \citenamefont {L{\"u}hmann}, \citenamefont {Heyl},
  \citenamefont {Budich}, \citenamefont {Mathey}, \citenamefont {Sengstock},\
  and\ \citenamefont {Weitenberg}}]{Weitenberg:nature}%
  \BibitemOpen
  \bibfield  {author} {\bibinfo {author} {\bibfnamefont {N.}~\bibnamefont
  {Fl{\"a}schner}}, \bibinfo {author} {\bibfnamefont {D.}~\bibnamefont
  {Vogel}}, \bibinfo {author} {\bibfnamefont {M.}~\bibnamefont {Tarnowski}},
  \bibinfo {author} {\bibfnamefont {B.~S.}\ \bibnamefont {Rem}}, \bibinfo
  {author} {\bibfnamefont {D.~S.}\ \bibnamefont {L{\"u}hmann}}, \bibinfo
  {author} {\bibfnamefont {M.}~\bibnamefont {Heyl}}, \bibinfo {author}
  {\bibfnamefont {J.~C.}\ \bibnamefont {Budich}}, \bibinfo {author}
  {\bibfnamefont {L.}~\bibnamefont {Mathey}}, \bibinfo {author} {\bibfnamefont
  {K.}~\bibnamefont {Sengstock}}, \ and\ \bibinfo {author} {\bibfnamefont
  {C.}~\bibnamefont {Weitenberg}},\ }\href {\doibase 10.1038/s41567-017-0013-8}
  {\bibfield  {journal} {\bibinfo  {journal} {Nat. Phys.}\ }\textbf {\bibinfo
  {volume} {14}},\ \bibinfo {pages} {265} (\bibinfo {year} {2018})}\BibitemShut
  {NoStop}%
\bibitem [{\citenamefont {Budich}\ and\ \citenamefont
  {Heyl}(2016)}]{PhysRevB.93.085416}%
  \BibitemOpen
  \bibfield  {author} {\bibinfo {author} {\bibfnamefont {J.~C.}\ \bibnamefont
  {Budich}}\ and\ \bibinfo {author} {\bibfnamefont {M.}~\bibnamefont {Heyl}},\
  }\href {\doibase 10.1103/PhysRevB.93.085416} {\bibfield  {journal} {\bibinfo
  {journal} {Phys. Rev. B}\ }\textbf {\bibinfo {volume} {93}},\ \bibinfo
  {pages} {085416} (\bibinfo {year} {2016})}\BibitemShut {NoStop}%
\bibitem [{\citenamefont {Vajna}\ and\ \citenamefont
  {D\'ora}(2015)}]{PhysRevB.91.155127}%
  \BibitemOpen
  \bibfield  {author} {\bibinfo {author} {\bibfnamefont {S.}~\bibnamefont
  {Vajna}}\ and\ \bibinfo {author} {\bibfnamefont {B.}~\bibnamefont {D\'ora}},\
  }\href {\doibase 10.1103/PhysRevB.91.155127} {\bibfield  {journal} {\bibinfo
  {journal} {Phys. Rev. B}\ }\textbf {\bibinfo {volume} {91}},\ \bibinfo
  {pages} {155127} (\bibinfo {year} {2015})}\BibitemShut {NoStop}%
\bibitem [{\citenamefont {Vajna}\ and\ \citenamefont
  {D\'ora}(2014)}]{PhysRevB.89.161105}%
  \BibitemOpen
  \bibfield  {author} {\bibinfo {author} {\bibfnamefont {S.}~\bibnamefont
  {Vajna}}\ and\ \bibinfo {author} {\bibfnamefont {B.}~\bibnamefont {D\'ora}},\
  }\href {\doibase 10.1103/PhysRevB.89.161105} {\bibfield  {journal} {\bibinfo
  {journal} {Phys. Rev. B}\ }\textbf {\bibinfo {volume} {89}},\ \bibinfo
  {pages} {161105(R)} (\bibinfo {year} {2014})}\BibitemShut {NoStop}%
\bibitem [{\citenamefont {Lang}\ \emph
  {et~al.}(2018{\natexlab{a}})\citenamefont {Lang}, \citenamefont {Frank},\
  and\ \citenamefont {Halimeh}}]{PhysRevB.97.174401}%
  \BibitemOpen
  \bibfield  {author} {\bibinfo {author} {\bibfnamefont {J.}~\bibnamefont
  {Lang}}, \bibinfo {author} {\bibfnamefont {B.}~\bibnamefont {Frank}}, \ and\
  \bibinfo {author} {\bibfnamefont {J.~C.}\ \bibnamefont {Halimeh}},\ }\href
  {\doibase 10.1103/PhysRevB.97.174401} {\bibfield  {journal} {\bibinfo
  {journal} {Phys. Rev. B}\ }\textbf {\bibinfo {volume} {97}},\ \bibinfo
  {pages} {174401} (\bibinfo {year} {2018}{\natexlab{a}})}\BibitemShut
  {NoStop}%
\bibitem [{\citenamefont {Lang}\ \emph
  {et~al.}(2018{\natexlab{b}})\citenamefont {Lang}, \citenamefont {Frank},\
  and\ \citenamefont {Halimeh}}]{PhysRevLett.121.130603}%
  \BibitemOpen
  \bibfield  {author} {\bibinfo {author} {\bibfnamefont {J.}~\bibnamefont
  {Lang}}, \bibinfo {author} {\bibfnamefont {B.}~\bibnamefont {Frank}}, \ and\
  \bibinfo {author} {\bibfnamefont {J.~C.}\ \bibnamefont {Halimeh}},\ }\href
  {\doibase 10.1103/PhysRevLett.121.130603} {\bibfield  {journal} {\bibinfo
  {journal} {Phys. Rev. Lett.}\ }\textbf {\bibinfo {volume} {121}},\ \bibinfo
  {pages} {130603} (\bibinfo {year} {2018}{\natexlab{b}})}\BibitemShut
  {NoStop}%
\bibitem [{\citenamefont {Weidinger}\ \emph {et~al.}(2017)\citenamefont
  {Weidinger}, \citenamefont {Heyl}, \citenamefont {Silva},\ and\ \citenamefont
  {Knap}}]{PhysRevB.96.134313}%
  \BibitemOpen
  \bibfield  {author} {\bibinfo {author} {\bibfnamefont {S.~A.}\ \bibnamefont
  {Weidinger}}, \bibinfo {author} {\bibfnamefont {M.}~\bibnamefont {Heyl}},
  \bibinfo {author} {\bibfnamefont {A.}~\bibnamefont {Silva}}, \ and\ \bibinfo
  {author} {\bibfnamefont {M.}~\bibnamefont {Knap}},\ }\href {\doibase
  10.1103/PhysRevB.96.134313} {\bibfield  {journal} {\bibinfo  {journal} {Phys.
  Rev. B}\ }\textbf {\bibinfo {volume} {96}},\ \bibinfo {pages} {134313}
  (\bibinfo {year} {2017})}\BibitemShut {NoStop}%
\bibitem [{\citenamefont {Halimeh}\ and\ \citenamefont
  {Zauner-Stauber}(2017)}]{PhysRevB.96.134427}%
  \BibitemOpen
  \bibfield  {author} {\bibinfo {author} {\bibfnamefont {J.~C.}\ \bibnamefont
  {Halimeh}}\ and\ \bibinfo {author} {\bibfnamefont {V.}~\bibnamefont
  {Zauner-Stauber}},\ }\href {\doibase 10.1103/PhysRevB.96.134427} {\bibfield
  {journal} {\bibinfo  {journal} {Phys. Rev. B}\ }\textbf {\bibinfo {volume}
  {96}},\ \bibinfo {pages} {134427} (\bibinfo {year} {2017})}\BibitemShut
  {NoStop}%
\bibitem [{\citenamefont {Zauner-Stauber}\ and\ \citenamefont
  {Halimeh}(2017)}]{PhysRevE.96.062118}%
  \BibitemOpen
  \bibfield  {author} {\bibinfo {author} {\bibfnamefont {V.}~\bibnamefont
  {Zauner-Stauber}}\ and\ \bibinfo {author} {\bibfnamefont {J.~C.}\
  \bibnamefont {Halimeh}},\ }\href {\doibase 10.1103/PhysRevE.96.062118}
  {\bibfield  {journal} {\bibinfo  {journal} {Phys. Rev. E}\ }\textbf {\bibinfo
  {volume} {96}},\ \bibinfo {pages} {062118} (\bibinfo {year}
  {2017})}\BibitemShut {NoStop}%
\bibitem [{\citenamefont {Homrighausen}\ \emph {et~al.}(2017)\citenamefont
  {Homrighausen}, \citenamefont {Abeling}, \citenamefont {Zauner-Stauber},\
  and\ \citenamefont {Halimeh}}]{PhysRevB.96.104436}%
  \BibitemOpen
  \bibfield  {author} {\bibinfo {author} {\bibfnamefont {I.}~\bibnamefont
  {Homrighausen}}, \bibinfo {author} {\bibfnamefont {N.~O.}\ \bibnamefont
  {Abeling}}, \bibinfo {author} {\bibfnamefont {V.}~\bibnamefont
  {Zauner-Stauber}}, \ and\ \bibinfo {author} {\bibfnamefont {J.~C.}\
  \bibnamefont {Halimeh}},\ }\href {\doibase 10.1103/PhysRevB.96.104436}
  {\bibfield  {journal} {\bibinfo  {journal} {Phys. Rev. B}\ }\textbf {\bibinfo
  {volume} {96}},\ \bibinfo {pages} {104436} (\bibinfo {year}
  {2017})}\BibitemShut {NoStop}%
\bibitem [{\citenamefont {Hashizume}\ \emph {et~al.}()\citenamefont
  {Hashizume}, \citenamefont {McCulloch},\ and\ \citenamefont
  {Halimeh}}]{Halimeh:preprint1}%
  \BibitemOpen
  \bibfield  {author} {\bibinfo {author} {\bibfnamefont {T.}~\bibnamefont
  {Hashizume}}, \bibinfo {author} {\bibfnamefont {I.~P.}\ \bibnamefont
  {McCulloch}}, \ and\ \bibinfo {author} {\bibfnamefont {J.~C.}\ \bibnamefont
  {Halimeh}},\ }\href@noop {} {\bibinfo  {journal} {arXiv:1811.09275}\
  }\BibitemShut {NoStop}%
\bibitem [{\citenamefont {Halimeh}\ \emph {et~al.}({\natexlab{a}})\citenamefont
  {Halimeh}, \citenamefont {Damme}, \citenamefont {Zauner-Stauber},\ and\
  \citenamefont {Vanderstraeten}}]{Halimeh:preprint2}%
  \BibitemOpen
\bibfield  {journal} {  }\bibfield  {author} {\bibinfo {author} {\bibfnamefont
  {J.~C.}\ \bibnamefont {Halimeh}}, \bibinfo {author} {\bibfnamefont {M.~V.}\
  \bibnamefont {Damme}}, \bibinfo {author} {\bibfnamefont {V.}~\bibnamefont
  {Zauner-Stauber}}, \ and\ \bibinfo {author} {\bibfnamefont {L.}~\bibnamefont
  {Vanderstraeten}},\ }\href@noop {} {\bibfield  {journal} {\bibinfo  {journal}
  {arXiv:1810.07187}\ } ({\natexlab{a}})}\BibitemShut {NoStop}%
\bibitem [{\citenamefont {Halimeh}\ \emph {et~al.}({\natexlab{b}})\citenamefont
  {Halimeh}, \citenamefont {Yegovtsev},\ and\ \citenamefont
  {Gurarie}}]{Halimeh:preprint3}%
  \BibitemOpen
  \bibfield  {author} {\bibinfo {author} {\bibfnamefont {J.~C.}\ \bibnamefont
  {Halimeh}}, \bibinfo {author} {\bibfnamefont {N.}~\bibnamefont {Yegovtsev}},
  \ and\ \bibinfo {author} {\bibfnamefont {V.}~\bibnamefont {Gurarie}},\
  }\href@noop {} {\bibfield  {journal} {\bibinfo  {journal} {arXiv:1903.03109}\
  } ({\natexlab{b}})}\BibitemShut {NoStop}%
\bibitem [{\citenamefont {Karrasch}\ and\ \citenamefont
  {Schuricht}(2013)}]{PhysRevB.87.195104}%
  \BibitemOpen
  \bibfield  {author} {\bibinfo {author} {\bibfnamefont {C.}~\bibnamefont
  {Karrasch}}\ and\ \bibinfo {author} {\bibfnamefont {D.}~\bibnamefont
  {Schuricht}},\ }\href {\doibase 10.1103/PhysRevB.87.195104} {\bibfield
  {journal} {\bibinfo  {journal} {Phys. Rev. B}\ }\textbf {\bibinfo {volume}
  {87}},\ \bibinfo {pages} {195104} (\bibinfo {year} {2013})}\BibitemShut
  {NoStop}%
\bibitem [{\citenamefont {Fogarty}\ \emph {et~al.}(2017)\citenamefont
  {Fogarty}, \citenamefont {Usui}, \citenamefont {Busch}, \citenamefont
  {Silva},\ and\ \citenamefont {Goold}}]{Fogarty_2017}%
  \BibitemOpen
  \bibfield  {author} {\bibinfo {author} {\bibfnamefont {T.}~\bibnamefont
  {Fogarty}}, \bibinfo {author} {\bibfnamefont {A.}~\bibnamefont {Usui}},
  \bibinfo {author} {\bibfnamefont {T.}~\bibnamefont {Busch}}, \bibinfo
  {author} {\bibfnamefont {A.}~\bibnamefont {Silva}}, \ and\ \bibinfo {author}
  {\bibfnamefont {J.}~\bibnamefont {Goold}},\ }\href {\doibase
  10.1088/1367-2630/aa8aff} {\bibfield  {journal} {\bibinfo  {journal} {New J.
  Phys.}\ }\textbf {\bibinfo {volume} {19}},\ \bibinfo {pages} {113018}
  (\bibinfo {year} {2017})}\BibitemShut {NoStop}%
\bibitem [{\citenamefont {Darmawan}\ and\ \citenamefont
  {Bartlett}(2010)}]{PhysRevA.82.012328}%
  \BibitemOpen
  \bibfield  {author} {\bibinfo {author} {\bibfnamefont {A.~S.}\ \bibnamefont
  {Darmawan}}\ and\ \bibinfo {author} {\bibfnamefont {S.~D.}\ \bibnamefont
  {Bartlett}},\ }\href {\doibase 10.1103/PhysRevA.82.012328} {\bibfield
  {journal} {\bibinfo  {journal} {Phys. Rev. A}\ }\textbf {\bibinfo {volume}
  {82}},\ \bibinfo {pages} {012328} (\bibinfo {year} {2010})}\BibitemShut
  {NoStop}%
\bibitem [{\citenamefont {Else}\ \emph {et~al.}(2012)\citenamefont {Else},
  \citenamefont {Schwarz}, \citenamefont {Bartlett},\ and\ \citenamefont
  {Doherty}}]{PhysRevLett.108.240505}%
  \BibitemOpen
  \bibfield  {author} {\bibinfo {author} {\bibfnamefont {D.~V.}\ \bibnamefont
  {Else}}, \bibinfo {author} {\bibfnamefont {I.}~\bibnamefont {Schwarz}},
  \bibinfo {author} {\bibfnamefont {S.~D.}\ \bibnamefont {Bartlett}}, \ and\
  \bibinfo {author} {\bibfnamefont {A.~C.}\ \bibnamefont {Doherty}},\ }\href
  {\doibase 10.1103/PhysRevLett.108.240505} {\bibfield  {journal} {\bibinfo
  {journal} {Phys. Rev. Lett.}\ }\textbf {\bibinfo {volume} {108}},\ \bibinfo
  {pages} {240505} (\bibinfo {year} {2012})}\BibitemShut {NoStop}%
\bibitem [{\citenamefont {McGinley}\ and\ \citenamefont
  {Cooper}(2019)}]{PhysRevB.99.075148}%
  \BibitemOpen
  \bibfield  {author} {\bibinfo {author} {\bibfnamefont {M.}~\bibnamefont
  {McGinley}}\ and\ \bibinfo {author} {\bibfnamefont {N.~R.}\ \bibnamefont
  {Cooper}},\ }\href {\doibase 10.1103/PhysRevB.99.075148} {\bibfield
  {journal} {\bibinfo  {journal} {Phys. Rev. B}\ }\textbf {\bibinfo {volume}
  {99}},\ \bibinfo {pages} {075148} (\bibinfo {year} {2019})}\BibitemShut
  {NoStop}%
\bibitem [{\citenamefont {Calvanese~Strinati}\ \emph
  {et~al.}(2016)\citenamefont {Calvanese~Strinati}, \citenamefont {Mazza},
  \citenamefont {Endres}, \citenamefont {Rossini},\ and\ \citenamefont
  {Fazio}}]{PhysRevB.94.024302}%
  \BibitemOpen
  \bibfield  {author} {\bibinfo {author} {\bibfnamefont {M.}~\bibnamefont
  {Calvanese~Strinati}}, \bibinfo {author} {\bibfnamefont {L.}~\bibnamefont
  {Mazza}}, \bibinfo {author} {\bibfnamefont {M.}~\bibnamefont {Endres}},
  \bibinfo {author} {\bibfnamefont {D.}~\bibnamefont {Rossini}}, \ and\
  \bibinfo {author} {\bibfnamefont {R.}~\bibnamefont {Fazio}},\ }\href
  {\doibase 10.1103/PhysRevB.94.024302} {\bibfield  {journal} {\bibinfo
  {journal} {Phys. Rev. B}\ }\textbf {\bibinfo {volume} {94}},\ \bibinfo
  {pages} {024302} (\bibinfo {year} {2016})}\BibitemShut {NoStop}%
\bibitem [{\citenamefont {Mazza}\ \emph {et~al.}(2014)\citenamefont {Mazza},
  \citenamefont {Rossini}, \citenamefont {Endres},\ and\ \citenamefont
  {Fazio}}]{PhysRevB.90.020301}%
  \BibitemOpen
  \bibfield  {author} {\bibinfo {author} {\bibfnamefont {L.}~\bibnamefont
  {Mazza}}, \bibinfo {author} {\bibfnamefont {D.}~\bibnamefont {Rossini}},
  \bibinfo {author} {\bibfnamefont {M.}~\bibnamefont {Endres}}, \ and\ \bibinfo
  {author} {\bibfnamefont {R.}~\bibnamefont {Fazio}},\ }\href {\doibase
  10.1103/PhysRevB.90.020301} {\bibfield  {journal} {\bibinfo  {journal} {Phys.
  Rev. B}\ }\textbf {\bibinfo {volume} {90}},\ \bibinfo {pages} {020301(R)}
  (\bibinfo {year} {2014})}\BibitemShut {NoStop}%
\bibitem [{\citenamefont {McGinley}\ and\ \citenamefont
  {Cooper}(2018)}]{PhysRevLett.121.090401}%
  \BibitemOpen
  \bibfield  {author} {\bibinfo {author} {\bibfnamefont {M.}~\bibnamefont
  {McGinley}}\ and\ \bibinfo {author} {\bibfnamefont {N.~R.}\ \bibnamefont
  {Cooper}},\ }\href {\doibase 10.1103/PhysRevLett.121.090401} {\bibfield
  {journal} {\bibinfo  {journal} {Phys. Rev. Lett.}\ }\textbf {\bibinfo
  {volume} {121}},\ \bibinfo {pages} {090401} (\bibinfo {year}
  {2018})}\BibitemShut {NoStop}%
\bibitem [{\citenamefont {Botet}\ \emph {et~al.}(1983)\citenamefont {Botet},
  \citenamefont {Jullien},\ and\ \citenamefont {Kolb}}]{PhysRevB.28.3914}%
  \BibitemOpen
  \bibfield  {author} {\bibinfo {author} {\bibfnamefont {R.}~\bibnamefont
  {Botet}}, \bibinfo {author} {\bibfnamefont {R.}~\bibnamefont {Jullien}}, \
  and\ \bibinfo {author} {\bibfnamefont {M.}~\bibnamefont {Kolb}},\ }\href
  {\doibase 10.1103/PhysRevB.28.3914} {\bibfield  {journal} {\bibinfo
  {journal} {Phys. Rev. B}\ }\textbf {\bibinfo {volume} {28}},\ \bibinfo
  {pages} {3914} (\bibinfo {year} {1983})}\BibitemShut {NoStop}%
\bibitem [{\citenamefont {Chen}\ \emph {et~al.}(2003)\citenamefont {Chen},
  \citenamefont {Hida},\ and\ \citenamefont {Sanctuary}}]{PhysRevB.67.104401}%
  \BibitemOpen
  \bibfield  {author} {\bibinfo {author} {\bibfnamefont {W.}~\bibnamefont
  {Chen}}, \bibinfo {author} {\bibfnamefont {K.}~\bibnamefont {Hida}}, \ and\
  \bibinfo {author} {\bibfnamefont {B.~C.}\ \bibnamefont {Sanctuary}},\ }\href
  {\doibase 10.1103/PhysRevB.67.104401} {\bibfield  {journal} {\bibinfo
  {journal} {Phys. Rev. B}\ }\textbf {\bibinfo {volume} {67}},\ \bibinfo
  {pages} {104401} (\bibinfo {year} {2003})}\BibitemShut {NoStop}%
\bibitem [{\citenamefont {Affleck}\ \emph {et~al.}(1987)\citenamefont
  {Affleck}, \citenamefont {Kennedy}, \citenamefont {Lieb},\ and\ \citenamefont
  {Tasaki}}]{PhysRevLett.59.799}%
  \BibitemOpen
  \bibfield  {author} {\bibinfo {author} {\bibfnamefont {I.}~\bibnamefont
  {Affleck}}, \bibinfo {author} {\bibfnamefont {T.}~\bibnamefont {Kennedy}},
  \bibinfo {author} {\bibfnamefont {E.~H.}\ \bibnamefont {Lieb}}, \ and\
  \bibinfo {author} {\bibfnamefont {H.}~\bibnamefont {Tasaki}},\ }\href
  {\doibase 10.1103/PhysRevLett.59.799} {\bibfield  {journal} {\bibinfo
  {journal} {Phys. Rev. Lett.}\ }\textbf {\bibinfo {volume} {59}},\ \bibinfo
  {pages} {799} (\bibinfo {year} {1987})}\BibitemShut {NoStop}%
\bibitem [{AKL()}]{AKLT}%
  \BibitemOpen
  \href@noop {} {}\bibinfo {note} {This special choice for the initial state
  does not restrict the validity of our calculations, since it belongs
  evidently to the Haldane phase and has simple matrix-product-state
  representation with bond dimension $M=2$, on the other hand it enables us to
  reach longer times, with high accuracy. Simulations with the ground state
  corresponding to $D=0$ and $\Delta=1$ does not give qualitatively different
  results.}\BibitemShut {Stop}%
\bibitem [{SI()}]{SI}%
  \BibitemOpen
  \href@noop {} {}\bibinfo {note} {See Supplemental Material at [URL will be
  inserted by publisher] for more numerical details and for the quench results
  with $\Delta<0$, which includes
  Refs.~[\onlinecite{PhysRevLett.107.070601,PhysRevB.94.165116,Hubig:review,Chiara_2006,Heyl_2018}].}\BibitemShut
  {Stop}%
\bibitem [{\citenamefont {Haegeman}\ \emph {et~al.}(2011)\citenamefont
  {Haegeman}, \citenamefont {Cirac}, \citenamefont {Osborne}, \citenamefont
  {Pi\ifmmode~\check{z}\else \v{z}\fi{}orn}, \citenamefont {Verschelde},\ and\
  \citenamefont {Verstraete}}]{PhysRevLett.107.070601}%
  \BibitemOpen
  \bibfield  {author} {\bibinfo {author} {\bibfnamefont {J.}~\bibnamefont
  {Haegeman}}, \bibinfo {author} {\bibfnamefont {J.~I.}\ \bibnamefont {Cirac}},
  \bibinfo {author} {\bibfnamefont {T.~J.}\ \bibnamefont {Osborne}}, \bibinfo
  {author} {\bibfnamefont {I.}~\bibnamefont {Pi\ifmmode~\check{z}\else
  \v{z}\fi{}orn}}, \bibinfo {author} {\bibfnamefont {H.}~\bibnamefont
  {Verschelde}}, \ and\ \bibinfo {author} {\bibfnamefont {F.}~\bibnamefont
  {Verstraete}},\ }\href {\doibase 10.1103/PhysRevLett.107.070601} {\bibfield
  {journal} {\bibinfo  {journal} {Phys. Rev. Lett.}\ }\textbf {\bibinfo
  {volume} {107}},\ \bibinfo {pages} {070601} (\bibinfo {year}
  {2011})}\BibitemShut {NoStop}%
\bibitem [{\citenamefont {Haegeman}\ \emph {et~al.}(2016)\citenamefont
  {Haegeman}, \citenamefont {Lubich}, \citenamefont {Oseledets}, \citenamefont
  {Vandereycken},\ and\ \citenamefont {Verstraete}}]{PhysRevB.94.165116}%
  \BibitemOpen
  \bibfield  {author} {\bibinfo {author} {\bibfnamefont {J.}~\bibnamefont
  {Haegeman}}, \bibinfo {author} {\bibfnamefont {C.}~\bibnamefont {Lubich}},
  \bibinfo {author} {\bibfnamefont {I.}~\bibnamefont {Oseledets}}, \bibinfo
  {author} {\bibfnamefont {B.}~\bibnamefont {Vandereycken}}, \ and\ \bibinfo
  {author} {\bibfnamefont {F.}~\bibnamefont {Verstraete}},\ }\href {\doibase
  10.1103/PhysRevB.94.165116} {\bibfield  {journal} {\bibinfo  {journal} {Phys.
  Rev. B}\ }\textbf {\bibinfo {volume} {94}},\ \bibinfo {pages} {165116}
  (\bibinfo {year} {2016})}\BibitemShut {NoStop}%
\bibitem [{\citenamefont {Paeckel}\ \emph {et~al.}()\citenamefont {Paeckel},
  \citenamefont {K\"ohler}, \citenamefont {Swoboda}, \citenamefont {Manmana},
  \citenamefont {Schollw\"ock},\ and\ \citenamefont {Hubig}}]{Hubig:review}%
  \BibitemOpen
  \bibfield  {author} {\bibinfo {author} {\bibfnamefont {S.}~\bibnamefont
  {Paeckel}}, \bibinfo {author} {\bibfnamefont {T.}~\bibnamefont {K\"ohler}},
  \bibinfo {author} {\bibfnamefont {A.}~\bibnamefont {Swoboda}}, \bibinfo
  {author} {\bibfnamefont {S.~R.}\ \bibnamefont {Manmana}}, \bibinfo {author}
  {\bibfnamefont {U.}~\bibnamefont {Schollw\"ock}}, \ and\ \bibinfo {author}
  {\bibfnamefont {C.}~\bibnamefont {Hubig}},\ }\href@noop {} {\bibinfo
  {journal} {arXiv:1901.05824}\ }\BibitemShut {NoStop}%
\bibitem [{\citenamefont {Piroli}\ \emph {et~al.}(2018)\citenamefont {Piroli},
  \citenamefont {Pozsgay},\ and\ \citenamefont {Vernier}}]{PIROLI2018454}%
  \BibitemOpen
\bibfield  {journal} {  }\bibfield  {author} {\bibinfo {author} {\bibfnamefont
  {L.}~\bibnamefont {Piroli}}, \bibinfo {author} {\bibfnamefont
  {B.}~\bibnamefont {Pozsgay}}, \ and\ \bibinfo {author} {\bibfnamefont
  {E.}~\bibnamefont {Vernier}},\ }\href {\doibase
  https://doi.org/10.1016/j.nuclphysb.2018.06.015} {\bibfield  {journal}
  {\bibinfo  {journal} {Nucl. Phys. B}\ }\textbf {\bibinfo {volume} {933}},\
  \bibinfo {pages} {454 } (\bibinfo {year} {2018})}\BibitemShut {NoStop}%
\bibitem [{\citenamefont {den Nijs}\ and\ \citenamefont
  {Rommelse}(1989)}]{PhysRevB.40.4709}%
  \BibitemOpen
  \bibfield  {author} {\bibinfo {author} {\bibfnamefont {M.}~\bibnamefont {den
  Nijs}}\ and\ \bibinfo {author} {\bibfnamefont {K.}~\bibnamefont {Rommelse}},\
  }\href {\doibase 10.1103/PhysRevB.40.4709} {\bibfield  {journal} {\bibinfo
  {journal} {Phys. Rev. B}\ }\textbf {\bibinfo {volume} {40}},\ \bibinfo
  {pages} {4709} (\bibinfo {year} {1989})}\BibitemShut {NoStop}%
\bibitem [{\citenamefont {Pollmann}\ \emph {et~al.}(2010)\citenamefont
  {Pollmann}, \citenamefont {Turner}, \citenamefont {Berg},\ and\ \citenamefont
  {Oshikawa}}]{Pollmann2010prb}%
  \BibitemOpen
  \bibfield  {author} {\bibinfo {author} {\bibfnamefont {F.}~\bibnamefont
  {Pollmann}}, \bibinfo {author} {\bibfnamefont {A.~M.}\ \bibnamefont
  {Turner}}, \bibinfo {author} {\bibfnamefont {E.}~\bibnamefont {Berg}}, \ and\
  \bibinfo {author} {\bibfnamefont {M.}~\bibnamefont {Oshikawa}},\ }\href@noop
  {} {\bibfield  {journal} {\bibinfo  {journal} {Phys. Rev. B}\ }\textbf
  {\bibinfo {volume} {81}},\ \bibinfo {pages} {064439} (\bibinfo {year}
  {2010})}\BibitemShut {NoStop}%
\bibitem [{\citenamefont {Pollmann}\ and\ \citenamefont
  {Turner}(2012)}]{Pollmann2012prb}%
  \BibitemOpen
  \bibfield  {author} {\bibinfo {author} {\bibfnamefont {F.}~\bibnamefont
  {Pollmann}}\ and\ \bibinfo {author} {\bibfnamefont {A.~M.}\ \bibnamefont
  {Turner}},\ }\href@noop {} {\bibfield  {journal} {\bibinfo  {journal} {Phys.
  Rev. B}\ }\textbf {\bibinfo {volume} {86}},\ \bibinfo {pages} {125441}
  (\bibinfo {year} {2012})}\BibitemShut {NoStop}%
\bibitem [{\citenamefont {Kl\"umper}\ \emph {et~al.}(1993)\citenamefont
  {Kl\"umper}, \citenamefont {Schadschneider},\ and\ \citenamefont
  {Zittartz}}]{Kl_mper_1993}%
  \BibitemOpen
  \bibfield  {author} {\bibinfo {author} {\bibfnamefont {A.}~\bibnamefont
  {Kl\"umper}}, \bibinfo {author} {\bibfnamefont {A.}~\bibnamefont
  {Schadschneider}}, \ and\ \bibinfo {author} {\bibfnamefont {J.}~\bibnamefont
  {Zittartz}},\ }\href {\doibase 10.1209/0295-5075/24/4/010} {\bibfield
  {journal} {\bibinfo  {journal} {Europhys. Lett. ({EPL})}\ }\textbf {\bibinfo
  {volume} {24}},\ \bibinfo {pages} {293} (\bibinfo {year} {1993})}\BibitemShut
  {NoStop}%
\bibitem [{\citenamefont {Eckstein}\ and\ \citenamefont
  {Werner}(2011)}]{PhysRevB.84.035122}%
  \BibitemOpen
  \bibfield  {author} {\bibinfo {author} {\bibfnamefont {M.}~\bibnamefont
  {Eckstein}}\ and\ \bibinfo {author} {\bibfnamefont {P.}~\bibnamefont
  {Werner}},\ }\href {\doibase 10.1103/PhysRevB.84.035122} {\bibfield
  {journal} {\bibinfo  {journal} {Phys. Rev. B}\ }\textbf {\bibinfo {volume}
  {84}},\ \bibinfo {pages} {035122} (\bibinfo {year} {2011})}\BibitemShut
  {NoStop}%
\bibitem [{\citenamefont {Werner}\ \emph {et~al.}(2012)\citenamefont {Werner},
  \citenamefont {Tsuji},\ and\ \citenamefont {Eckstein}}]{PhysRevB.86.205101}%
  \BibitemOpen
  \bibfield  {author} {\bibinfo {author} {\bibfnamefont {P.}~\bibnamefont
  {Werner}}, \bibinfo {author} {\bibfnamefont {N.}~\bibnamefont {Tsuji}}, \
  and\ \bibinfo {author} {\bibfnamefont {M.}~\bibnamefont {Eckstein}},\ }\href
  {\doibase 10.1103/PhysRevB.86.205101} {\bibfield  {journal} {\bibinfo
  {journal} {Phys. Rev. B}\ }\textbf {\bibinfo {volume} {86}},\ \bibinfo
  {pages} {205101} (\bibinfo {year} {2012})}\BibitemShut {NoStop}%
\bibitem [{\citenamefont {Legeza}\ and\ \citenamefont
  {S\'olyom}(2003)}]{legeza2003b}%
  \BibitemOpen
  \bibfield  {author} {\bibinfo {author} {\bibfnamefont {{\"O}.}~\bibnamefont
  {Legeza}}\ and\ \bibinfo {author} {\bibfnamefont {J.}~\bibnamefont
  {S\'olyom}},\ }\href@noop {} {\bibfield  {journal} {\bibinfo  {journal}
  {Phys. Rev. B}\ }\textbf {\bibinfo {volume} {68}},\ \bibinfo {pages} {195116}
  (\bibinfo {year} {2003})}\BibitemShut {NoStop}%
\bibitem [{\citenamefont {Vidal}\ \emph {et~al.}(2003)\citenamefont {Vidal},
  \citenamefont {Latorre}, \citenamefont {Rico},\ and\ \citenamefont
  {Kitaev}}]{vidallatorre03}%
  \BibitemOpen
  \bibfield  {author} {\bibinfo {author} {\bibfnamefont {G.}~\bibnamefont
  {Vidal}}, \bibinfo {author} {\bibfnamefont {J.~I.}\ \bibnamefont {Latorre}},
  \bibinfo {author} {\bibfnamefont {E.}~\bibnamefont {Rico}}, \ and\ \bibinfo
  {author} {\bibfnamefont {A.}~\bibnamefont {Kitaev}},\ }\href@noop {}
  {\bibfield  {journal} {\bibinfo  {journal} {Phys. Rev. Lett.}\ }\textbf
  {\bibinfo {volume} {90}},\ \bibinfo {pages} {227902} (\bibinfo {year}
  {2003})}\BibitemShut {NoStop}%
\bibitem [{\citenamefont {Calabrese}\ and\ \citenamefont
  {Cardy}(2004)}]{calabrese04}%
  \BibitemOpen
  \bibfield  {author} {\bibinfo {author} {\bibfnamefont {P.}~\bibnamefont
  {Calabrese}}\ and\ \bibinfo {author} {\bibfnamefont {J.}~\bibnamefont
  {Cardy}},\ }\href@noop {} {\bibfield  {journal} {\bibinfo  {journal} {J.
  Stat. Mech.}\ }\textbf {\bibinfo {volume} {2004}},\ \bibinfo {pages} {P06002}
  (\bibinfo {year} {2004})}\BibitemShut {NoStop}%
\bibitem [{\citenamefont {Rissler}\ \emph
  {et~al.}(2006{\natexlab{a}})\citenamefont {Rissler}, \citenamefont {Noack},\
  and\ \citenamefont {White}}]{rissler2006}%
  \BibitemOpen
  \bibfield  {author} {\bibinfo {author} {\bibfnamefont {J.}~\bibnamefont
  {Rissler}}, \bibinfo {author} {\bibfnamefont {R.~M.}\ \bibnamefont {Noack}},
  \ and\ \bibinfo {author} {\bibfnamefont {S.~R.}\ \bibnamefont {White}},\
  }\href@noop {} {\bibfield  {journal} {\bibinfo  {journal} {Chem. Phys.}\
  }\textbf {\bibinfo {volume} {323}},\ \bibinfo {pages} {519 } (\bibinfo {year}
  {2006}{\natexlab{a}})}\BibitemShut {NoStop}%
\bibitem [{\citenamefont {Legeza}\ and\ \citenamefont
  {S\'olyom}(2006)}]{legeza2006}%
  \BibitemOpen
  \bibfield  {author} {\bibinfo {author} {\bibfnamefont {{\"O}.}~\bibnamefont
  {Legeza}}\ and\ \bibinfo {author} {\bibfnamefont {J.}~\bibnamefont
  {S\'olyom}},\ }\href@noop {} {\bibfield  {journal} {\bibinfo  {journal}
  {Phys. Rev. Lett.}\ }\textbf {\bibinfo {volume} {96}},\ \bibinfo {pages}
  {116401} (\bibinfo {year} {2006})}\BibitemShut {NoStop}%
\bibitem [{\citenamefont {Amico}\ \emph {et~al.}(2008)\citenamefont {Amico},
  \citenamefont {Fazio}, \citenamefont {Osterloh},\ and\ \citenamefont
  {Vedral}}]{luigi2008}%
  \BibitemOpen
  \bibfield  {author} {\bibinfo {author} {\bibfnamefont {L.}~\bibnamefont
  {Amico}}, \bibinfo {author} {\bibfnamefont {R.}~\bibnamefont {Fazio}},
  \bibinfo {author} {\bibfnamefont {A.}~\bibnamefont {Osterloh}}, \ and\
  \bibinfo {author} {\bibfnamefont {V.}~\bibnamefont {Vedral}},\ }\href@noop {}
  {\bibfield  {journal} {\bibinfo  {journal} {Rev. Mod. Phys.}\ }\textbf
  {\bibinfo {volume} {80}},\ \bibinfo {pages} {517} (\bibinfo {year}
  {2008})}\BibitemShut {NoStop}%
\bibitem [{\citenamefont {Gu}\ \emph {et~al.}(2004)\citenamefont {Gu},
  \citenamefont {Deng}, \citenamefont {Li},\ and\ \citenamefont
  {Lin}}]{gu:prl2004}%
  \BibitemOpen
  \bibfield  {author} {\bibinfo {author} {\bibfnamefont {S.-J.}\ \bibnamefont
  {Gu}}, \bibinfo {author} {\bibfnamefont {S.-S.}\ \bibnamefont {Deng}},
  \bibinfo {author} {\bibfnamefont {Y.-Q.}\ \bibnamefont {Li}}, \ and\ \bibinfo
  {author} {\bibfnamefont {H.-Q.}\ \bibnamefont {Lin}},\ }\href@noop {}
  {\bibfield  {journal} {\bibinfo  {journal} {Phys. Rev. Lett.}\ }\textbf
  {\bibinfo {volume} {93}},\ \bibinfo {pages} {086402} (\bibinfo {year}
  {2004})}\BibitemShut {NoStop}%
\bibitem [{\citenamefont {Wu}\ \emph {et~al.}(2004)\citenamefont {Wu},
  \citenamefont {Sarandy},\ and\ \citenamefont {Lidar}}]{wu:prl2004}%
  \BibitemOpen
  \bibfield  {author} {\bibinfo {author} {\bibfnamefont {L.-A.}\ \bibnamefont
  {Wu}}, \bibinfo {author} {\bibfnamefont {M.~S.}\ \bibnamefont {Sarandy}}, \
  and\ \bibinfo {author} {\bibfnamefont {D.~A.}\ \bibnamefont {Lidar}},\
  }\href@noop {} {\bibfield  {journal} {\bibinfo  {journal} {Phys. Rev. Lett.}\
  }\textbf {\bibinfo {volume} {93}},\ \bibinfo {pages} {250404} (\bibinfo
  {year} {2004})}\BibitemShut {NoStop}%
\bibitem [{\citenamefont {Yang}(2005)}]{yang:pra2005}%
  \BibitemOpen
  \bibfield  {author} {\bibinfo {author} {\bibfnamefont {M.-F.}\ \bibnamefont
  {Yang}},\ }\href@noop {} {\bibfield  {journal} {\bibinfo  {journal} {Phys.
  Rev. A}\ }\textbf {\bibinfo {volume} {71}},\ \bibinfo {pages} {030302(R)}
  (\bibinfo {year} {2005})}\BibitemShut {NoStop}%
\bibitem [{\citenamefont {Deng}\ \emph {et~al.}(2006)\citenamefont {Deng},
  \citenamefont {Gu},\ and\ \citenamefont {Lin}}]{deng:prb2006}%
  \BibitemOpen
  \bibfield  {author} {\bibinfo {author} {\bibfnamefont {S.-S.}\ \bibnamefont
  {Deng}}, \bibinfo {author} {\bibfnamefont {S.-J.}\ \bibnamefont {Gu}}, \ and\
  \bibinfo {author} {\bibfnamefont {H.-Q.}\ \bibnamefont {Lin}},\ }\href@noop
  {} {\bibfield  {journal} {\bibinfo  {journal} {Phys. Rev. B}\ }\textbf
  {\bibinfo {volume} {74}},\ \bibinfo {pages} {045103} (\bibinfo {year}
  {2006})}\BibitemShut {NoStop}%
\bibitem [{\citenamefont {Barcza}\ \emph {et~al.}(2011)\citenamefont {Barcza},
  \citenamefont {Legeza}, \citenamefont {Marti},\ and\ \citenamefont
  {Reiher}}]{PhysRevA.83.012508}%
  \BibitemOpen
  \bibfield  {author} {\bibinfo {author} {\bibfnamefont {G.}~\bibnamefont
  {Barcza}}, \bibinfo {author} {\bibfnamefont {{\"O}.}~\bibnamefont {Legeza}},
  \bibinfo {author} {\bibfnamefont {K.~H.}\ \bibnamefont {Marti}}, \ and\
  \bibinfo {author} {\bibfnamefont {M.}~\bibnamefont {Reiher}},\ }\href
  {\doibase 10.1103/PhysRevA.83.012508} {\bibfield  {journal} {\bibinfo
  {journal} {Phys. Rev. A}\ }\textbf {\bibinfo {volume} {83}},\ \bibinfo
  {pages} {012508} (\bibinfo {year} {2011})}\BibitemShut {NoStop}%
\bibitem [{\citenamefont {Rissler}\ \emph
  {et~al.}(2006{\natexlab{b}})\citenamefont {Rissler}, \citenamefont {Noack},\
  and\ \citenamefont {White}}]{RISSLER2006519}%
  \BibitemOpen
  \bibfield  {author} {\bibinfo {author} {\bibfnamefont {J.}~\bibnamefont
  {Rissler}}, \bibinfo {author} {\bibfnamefont {R.~M.}\ \bibnamefont {Noack}},
  \ and\ \bibinfo {author} {\bibfnamefont {S.~R.}\ \bibnamefont {White}},\
  }\href {\doibase https://doi.org/10.1016/j.chemphys.2005.10.018} {\bibfield
  {journal} {\bibinfo  {journal} {Chemical Physics}\ }\textbf {\bibinfo
  {volume} {323}},\ \bibinfo {pages} {519 } (\bibinfo {year}
  {2006}{\natexlab{b}})}\BibitemShut {NoStop}%
\bibitem [{\citenamefont {Schultz}\ \emph {et~al.}(1964)\citenamefont
  {Schultz}, \citenamefont {Mattis},\ and\ \citenamefont
  {Lieb}}]{RevModPhys.36.856}%
  \BibitemOpen
  \bibfield  {author} {\bibinfo {author} {\bibfnamefont {T.~D.}\ \bibnamefont
  {Schultz}}, \bibinfo {author} {\bibfnamefont {D.~C.}\ \bibnamefont {Mattis}},
  \ and\ \bibinfo {author} {\bibfnamefont {E.~H.}\ \bibnamefont {Lieb}},\
  }\href {\doibase 10.1103/RevModPhys.36.856} {\bibfield  {journal} {\bibinfo
  {journal} {Rev. Mod. Phys.}\ }\textbf {\bibinfo {volume} {36}},\ \bibinfo
  {pages} {856} (\bibinfo {year} {1964})}\BibitemShut {NoStop}%
\bibitem [{\citenamefont {Lieb}\ and\ \citenamefont
  {Robinson}(1972)}]{Lieb1972}%
  \BibitemOpen
  \bibfield  {author} {\bibinfo {author} {\bibfnamefont {E.~H.}\ \bibnamefont
  {Lieb}}\ and\ \bibinfo {author} {\bibfnamefont {D.~W.}\ \bibnamefont
  {Robinson}},\ }\href {\doibase 10.1007/BF01645779} {\bibfield  {journal}
  {\bibinfo  {journal} {Commun. Math. Phys.}\ }\textbf {\bibinfo {volume}
  {28}},\ \bibinfo {pages} {251} (\bibinfo {year} {1972})}\BibitemShut
  {NoStop}%
\bibitem [{\citenamefont {Hilker}\ \emph {et~al.}(2017)\citenamefont {Hilker},
  \citenamefont {Salomon}, \citenamefont {Grusdt}, \citenamefont {Omran},
  \citenamefont {Boll}, \citenamefont {Demler}, \citenamefont {Bloch},\ and\
  \citenamefont {Gross}}]{Hilker484}%
  \BibitemOpen
  \bibfield  {author} {\bibinfo {author} {\bibfnamefont {T.~A.}\ \bibnamefont
  {Hilker}}, \bibinfo {author} {\bibfnamefont {G.}~\bibnamefont {Salomon}},
  \bibinfo {author} {\bibfnamefont {F.}~\bibnamefont {Grusdt}}, \bibinfo
  {author} {\bibfnamefont {A.}~\bibnamefont {Omran}}, \bibinfo {author}
  {\bibfnamefont {M.}~\bibnamefont {Boll}}, \bibinfo {author} {\bibfnamefont
  {E.}~\bibnamefont {Demler}}, \bibinfo {author} {\bibfnamefont
  {I.}~\bibnamefont {Bloch}}, \ and\ \bibinfo {author} {\bibfnamefont
  {C.}~\bibnamefont {Gross}},\ }\href {\doibase 10.1126/science.aam8990}
  {\bibfield  {journal} {\bibinfo  {journal} {Science}\ }\textbf {\bibinfo
  {volume} {357}},\ \bibinfo {pages} {484} (\bibinfo {year}
  {2017})}\BibitemShut {NoStop}%
\bibitem [{\citenamefont {Senko}\ \emph {et~al.}(2015)\citenamefont {Senko},
  \citenamefont {Richerme}, \citenamefont {Smith}, \citenamefont {Lee},
  \citenamefont {Cohen}, \citenamefont {Retzker},\ and\ \citenamefont
  {Monroe}}]{PhysRevX.5.021026}%
  \BibitemOpen
  \bibfield  {author} {\bibinfo {author} {\bibfnamefont {C.}~\bibnamefont
  {Senko}}, \bibinfo {author} {\bibfnamefont {P.}~\bibnamefont {Richerme}},
  \bibinfo {author} {\bibfnamefont {J.}~\bibnamefont {Smith}}, \bibinfo
  {author} {\bibfnamefont {A.}~\bibnamefont {Lee}}, \bibinfo {author}
  {\bibfnamefont {I.}~\bibnamefont {Cohen}}, \bibinfo {author} {\bibfnamefont
  {A.}~\bibnamefont {Retzker}}, \ and\ \bibinfo {author} {\bibfnamefont
  {C.}~\bibnamefont {Monroe}},\ }\href {\doibase 10.1103/PhysRevX.5.021026}
  {\bibfield  {journal} {\bibinfo  {journal} {Phys. Rev. X}\ }\textbf {\bibinfo
  {volume} {5}},\ \bibinfo {pages} {021026} (\bibinfo {year}
  {2015})}\BibitemShut {NoStop}%
\bibitem [{\citenamefont {Chiara}\ \emph {et~al.}(2006)\citenamefont {Chiara},
  \citenamefont {Montangero}, \citenamefont {Calabrese},\ and\ \citenamefont
  {Fazio}}]{Chiara_2006}%
  \BibitemOpen
  \bibfield  {author} {\bibinfo {author} {\bibfnamefont {G.~D.}\ \bibnamefont
  {Chiara}}, \bibinfo {author} {\bibfnamefont {S.}~\bibnamefont {Montangero}},
  \bibinfo {author} {\bibfnamefont {P.}~\bibnamefont {Calabrese}}, \ and\
  \bibinfo {author} {\bibfnamefont {R.}~\bibnamefont {Fazio}},\ }\href
  {\doibase 10.1088/1742-5468/2006/03/p03001} {\bibfield  {journal} {\bibinfo
  {journal} {J. Stat. Mech.: Theory and Exp.}\ }\textbf {\bibinfo {volume}
  {2006}},\ \bibinfo {pages} {P03001} (\bibinfo {year} {2006})}\BibitemShut
  {NoStop}%
\end{thebibliography}%


\begin{thebibliography}{5}%
\makeatletter
\providecommand \@ifxundefined [1]{%
 \@ifx{#1\undefined}
}%
\providecommand \@ifnum [1]{%
 \ifnum #1\expandafter \@firstoftwo
 \else \expandafter \@secondoftwo
 \fi
}%
\providecommand \@ifx [1]{%
 \ifx #1\expandafter \@firstoftwo
 \else \expandafter \@secondoftwo
 \fi
}%
\providecommand \natexlab [1]{#1}%
\providecommand \enquote  [1]{``#1''}%
\providecommand \bibnamefont  [1]{#1}%
\providecommand \bibfnamefont [1]{#1}%
\providecommand \citenamefont [1]{#1}%
\providecommand \href@noop [0]{\@secondoftwo}%
\providecommand \href [0]{\begingroup \@sanitize@url \@href}%
\providecommand \@href[1]{\@@startlink{#1}\@@href}%
\providecommand \@@href[1]{\endgroup#1\@@endlink}%
\providecommand \@sanitize@url [0]{\catcode `\\12\catcode `\$12\catcode
  `\&12\catcode `\#12\catcode `\^12\catcode `\_12\catcode `\%12\relax}%
\providecommand \@@startlink[1]{}%
\providecommand \@@endlink[0]{}%
\providecommand \url  [0]{\begingroup\@sanitize@url \@url }%
\providecommand \@url [1]{\endgroup\@href {#1}{\urlprefix }}%
\providecommand \urlprefix  [0]{URL }%
\providecommand \Eprint [0]{\href }%
\providecommand \doibase [0]{http://dx.doi.org/}%
\providecommand \selectlanguage [0]{\@gobble}%
\providecommand \bibinfo  [0]{\@secondoftwo}%
\providecommand \bibfield  [0]{\@secondoftwo}%
\providecommand \translation [1]{[#1]}%
\providecommand \BibitemOpen [0]{}%
\providecommand \bibitemStop [0]{}%
\providecommand \bibitemNoStop [0]{.\EOS\space}%
\providecommand \EOS [0]{\spacefactor3000\relax}%
\providecommand \BibitemShut  [1]{\csname bibitem#1\endcsname}%
\let\auto@bib@innerbib\@empty
%</preamble>
\bibitem [{\citenamefont {Haegeman}\ \emph {et~al.}(2011)\citenamefont
  {Haegeman}, \citenamefont {Cirac}, \citenamefont {Osborne}, \citenamefont
  {Pi\ifmmode~\check{z}\else \v{z}\fi{}orn}, \citenamefont {Verschelde},\ and\
  \citenamefont {Verstraete}}]{PhysRevLett.107.070601}%
  \BibitemOpen
  \bibfield  {author} {\bibinfo {author} {\bibfnamefont {J.}~\bibnamefont
  {Haegeman}}, \bibinfo {author} {\bibfnamefont {J.~I.}\ \bibnamefont {Cirac}},
  \bibinfo {author} {\bibfnamefont {T.~J.}\ \bibnamefont {Osborne}}, \bibinfo
  {author} {\bibfnamefont {I.}~\bibnamefont {Pi\ifmmode~\check{z}\else
  \v{z}\fi{}orn}}, \bibinfo {author} {\bibfnamefont {H.}~\bibnamefont
  {Verschelde}}, \ and\ \bibinfo {author} {\bibfnamefont {F.}~\bibnamefont
  {Verstraete}},\ }\href {\doibase 10.1103/PhysRevLett.107.070601} {\bibfield
  {journal} {\bibinfo  {journal} {Phys. Rev. Lett.}\ }\textbf {\bibinfo
  {volume} {107}},\ \bibinfo {pages} {070601} (\bibinfo {year}
  {2011})}\BibitemShut {NoStop}%
\bibitem [{\citenamefont {Haegeman}\ \emph {et~al.}(2016)\citenamefont
  {Haegeman}, \citenamefont {Lubich}, \citenamefont {Oseledets}, \citenamefont
  {Vandereycken},\ and\ \citenamefont {Verstraete}}]{PhysRevB.94.165116}%
  \BibitemOpen
  \bibfield  {author} {\bibinfo {author} {\bibfnamefont {J.}~\bibnamefont
  {Haegeman}}, \bibinfo {author} {\bibfnamefont {C.}~\bibnamefont {Lubich}},
  \bibinfo {author} {\bibfnamefont {I.}~\bibnamefont {Oseledets}}, \bibinfo
  {author} {\bibfnamefont {B.}~\bibnamefont {Vandereycken}}, \ and\ \bibinfo
  {author} {\bibfnamefont {F.}~\bibnamefont {Verstraete}},\ }\href {\doibase
  10.1103/PhysRevB.94.165116} {\bibfield  {journal} {\bibinfo  {journal} {Phys.
  Rev. B}\ }\textbf {\bibinfo {volume} {94}},\ \bibinfo {pages} {165116}
  (\bibinfo {year} {2016})}\BibitemShut {NoStop}%
\bibitem [{\citenamefont {Paeckel}\ \emph {et~al.}()\citenamefont {Paeckel},
  \citenamefont {K\"ohler}, \citenamefont {Swoboda}, \citenamefont {Manmana},
  \citenamefont {Schollw\"ock},\ and\ \citenamefont {Hubig}}]{Hubig:review}%
  \BibitemOpen
  \bibfield  {author} {\bibinfo {author} {\bibfnamefont {S.}~\bibnamefont
  {Paeckel}}, \bibinfo {author} {\bibfnamefont {T.}~\bibnamefont {K\"ohler}},
  \bibinfo {author} {\bibfnamefont {A.}~\bibnamefont {Swoboda}}, \bibinfo
  {author} {\bibfnamefont {S.~R.}\ \bibnamefont {Manmana}}, \bibinfo {author}
  {\bibfnamefont {U.}~\bibnamefont {Schollw\"ock}}, \ and\ \bibinfo {author}
  {\bibfnamefont {C.}~\bibnamefont {Hubig}},\ }\href@noop {} {\bibinfo
  {journal} {arXiv:1901.05824}\ }\BibitemShut {NoStop}%
\bibitem [{\citenamefont {Chiara}\ \emph {et~al.}(2006)\citenamefont {Chiara},
  \citenamefont {Montangero}, \citenamefont {Calabrese},\ and\ \citenamefont
  {Fazio}}]{Chiara_2006}%
  \BibitemOpen
\bibfield  {journal} {  }\bibfield  {author} {\bibinfo {author} {\bibfnamefont
  {G.~D.}\ \bibnamefont {Chiara}}, \bibinfo {author} {\bibfnamefont
  {S.}~\bibnamefont {Montangero}}, \bibinfo {author} {\bibfnamefont
  {P.}~\bibnamefont {Calabrese}}, \ and\ \bibinfo {author} {\bibfnamefont
  {R.}~\bibnamefont {Fazio}},\ }\href {\doibase
  10.1088/1742-5468/2006/03/p03001} {\bibfield  {journal} {\bibinfo  {journal}
  {J. Stat. Mech.: Theory and Exp.}\ }\textbf {\bibinfo {volume} {2006}},\
  \bibinfo {pages} {P03001} (\bibinfo {year} {2006})}\BibitemShut {NoStop}%
\bibitem [{\citenamefont {Heyl}(2018)}]{Heyl_2018}%
  \BibitemOpen
  \bibfield  {author} {\bibinfo {author} {\bibfnamefont {M.}~\bibnamefont
  {Heyl}},\ }\href {\doibase 10.1088/1361-6633/aaaf9a} {\bibfield  {journal}
  {\bibinfo  {journal} {Rep. Prog. Phys.}\ }\textbf {\bibinfo {volume} {81}},\
  \bibinfo {pages} {054001} (\bibinfo {year} {2018})}\BibitemShut {NoStop}%
\end{thebibliography}%

\end{document}

\title{Supplemental Material for "Dynamical topological quantum phase transitions in nonintegrable models"}

\author{I. Hagym\'asi}
\affiliation{Department of Physics,
Arnold Sommerfeld Center for Theoretical Physics (ASC),
Fakult\"{a}t f\"{u}r Physik, Ludwig-Maximilians-Universit\"{a}t M\"{u}nchen,
D-80333 M\"{u}nchen, Germany}
\affiliation{Munich Center for Quantum Science and Technology (MCQST), Schellingstr. 4, D-80799 M\"unchen, Germany}
\affiliation{Strongly Correlated Systems "Lend\"ulet" Research Group, Institute for Solid State
Physics and Optics, MTA Wigner Research Centre for Physics, Budapest H-1525 P.O. Box 49, Hungary
}

\author{C. Hubig}
\affiliation{Max-Planck-Institut f\"ur Quantenoptik,
Hans-Kopfermann-Strasse 1, 85748 Garching, Germany}

\author{\"O. Legeza}
\affiliation{Strongly Correlated Systems "Lend\"ulet" Research Group, Institute for Solid State
Physics and Optics, MTA Wigner Research Centre for Physics, Budapest H-1525 P.O. Box 49, Hungary
}

\author{U. Schollw\"ock}
\affiliation{Department of Physics,
Arnold Sommerfeld Center for Theoretical Physics (ASC),
Fakult\"{a}t f\"{u}r Physik, Ludwig-Maximilians-Universit\"{a}t M\"{u}nchen,
D-80333 M\"{u}nchen, Germany}
\affiliation{Munich Center for Quantum Science and Technology (MCQST), Schellingstr. 4, D-80799 M\"unchen, Germany}

\maketitle

\section{Details of the numerical calculations}
The time evolution is performed using the two-site 
time-dependent 
variational principle (TDVP) 
method \cite{PhysRevLett.107.070601,PhysRevB.94.165116,Hubig:review}. Since the Hamiltonian contains only nearest-neighbor interactions, no additional errors 
emerge during 
the projection in the TDVP method, and the truncation error as well as the error coming from the 
inexact 
exponentialization is kept under control and smaller than $10^{-10}$. Such a small truncation error is 
necessary in 
order to calculate overlaps with time-evolved wave functions accurately. In 
our simulation we use the $U(1)$ 
symmetry and the maximal bond dimension reached is $\sim 6000$.  We consider chain lengths 
from $L=30$ 
to $L=120$ to account for finite-size effects, and we show results for $L=80$ unless mentioned 
otherwise. The 
greatest obstacle which limits the study of the dynamics in interacting systems is the linear 
growth of the 
entanglement entropy in time \cite{Chiara_2006}. Fortunately, from the point of 
view of DQPTs the interesting time 
scale is usually in 
the order of the inverse coupling in the system \cite{Heyl_2018} ($J^{-1}$ in our 
case), which makes the problem 
accurately 
treatable with matrix-product-state methods, before significant truncation errors set in.
\section{Quench to the XY and ferromagnetic phases}
\begin{figure}[!t]
\vspace{0.5cm}
\includegraphics[scale=0.6]{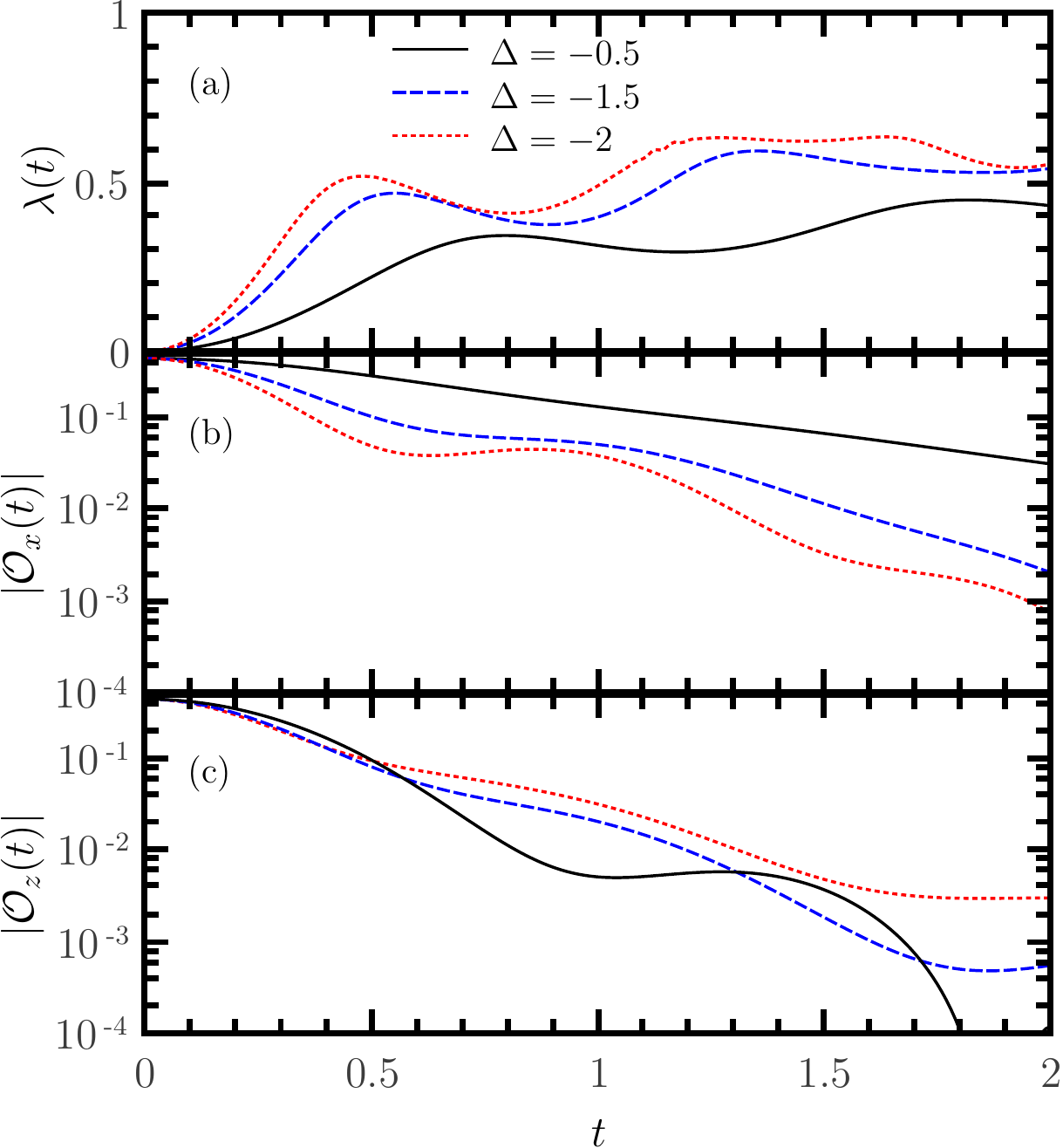}
\caption{(a) The rate function for various values of $\Delta$ and $D=0$. Panels (b) and (c) show 
the 
$x$- and $z$-component of the string order parameter for the same values of $\Delta$ and $D=0$ as 
in panel (a). The chain length is $L=80$ in all cases.}
\label{fig:quench-Delta-neg}
\end{figure}

We present here the results for  the quenches with $\Delta<0$, which corresponds to probing the spectra of the 
XY and  
the ferromagnetic phases. The results are shown in Fig.~\ref{fig:quench-Delta-neg}.
We find that independent of whether we quench to the XY or ferromagnetic phase, the rate 
function 
behaves always analytically in our time window. The string order parameters also obey this behavior,
 they decay in time with some superimposed oscillation. It is interesting to 
mention that 
$\mathcal{O}^x(t)$ decays much slower in time in contrast to $\mathcal{O}^z(t)$. 
We
checked quenches to other regions of the XY phase, and obtained qualitatively the same results as presented in 
Fig.~\ref{fig:quench-Delta-neg}.  
\section{Quench with inversion-symmetry breaking}
We address also the case of inversion-symmetry breaking in the XXZ Hamiltonian:
\begin{equation}
\begin{split}
\mathcal{H}=&\sum_{i=1}^{L-1} [J(S^x_iS^x_{i+1}+S^y_iS^y_{i+1}) + \Delta S^z_iS^z_{i+1} ] + \\
&B\sum_{i=1}^{L-1} [(S^x_iS^y_{i+1})^2-(S^y_iS^x_{i+1})^2],
\end{split}
\end{equation}
where the term proportional to $B$ describes the strength of the symmetry breaking. In this case the Hamiltonian still has the dihedral symmetry that protects the Haldane phase and supports the presence of string order. In Fig.~\ref{fig:quench-Delta-inv} we show the results for the quench to the N\'eel phase with nonzero value of $B$ starting from the AKLT state. The rate function and string order parameters with finite $B$ exhibit qualitatively the same results that are presented in the main text.
\begin{figure}[!t]
\vspace{0.5cm}
\includegraphics[scale=0.6]{quench-Delta-inv.pdf}
\caption{(a) The rate function for the quench parameters $\Delta=3$ and $B=0.5$. Panels (b) and (c) show 
the 
$x$- and $z$-component of the string order parameter for the same set of parameters as 
in panel (a). The chain length is $L=80$ in all cases and the truncation error was set to $10^{-7}$.}
\label{fig:quench-Delta-inv}
\end{figure}
\bibliography{tki_refs.bib}